\documentclass[aps,twocolumn,showpacs,groupedaddress,superscriptaddress,amsmath,amssymb]{revtex4}

\usepackage{graphicx}
\usepackage{dcolumn}
\usepackage{bm}
\usepackage{epsfig,amsmath}

\def\vec#1{{\bf #1}}

\begin{document}

\title{
(In)commensurability, scaling and multiplicity of friction in nanocrystals and application to gold nanocrystals on graphite
}

\author{Astrid~S. de Wijn\footnote{e-mail:A.S.deWijn@science.ru.nl}}
\affiliation{Department of Physics, Stockholm University, 106 91 Stockholm, Sweden}
\affiliation{Radboud University Nijmegen, Institute for Molecules and Materials, Heyendaalseweg 135, 6525AJ Nijmegen, the Netherlands}

\begin{abstract}

The scaling of friction with the contact size $A$ and (in)commensurabilty of nanoscopic and mesoscopic crystals on a regular substrate are investigated analytically for triangular nanocrystals on hexagonal substrates.
The crystals are assumed to be stiff, but not completely rigid.
Commensurate and incommensurate configurations are identified systematically.
It is shown that 
three distinct friction branches coexist, an incommensurate one that does not scale with the contact size ($A^0$) and two commensurate ones which scale differently (with $A^{1/2}$ and $A$) and are associated with various combinations of commensurate and incommensurate lattice parameters and orientations.
This coexistence is a direct consequence of the two-dimensional nature of the contact layer, and such multiplicity exists in all geometries consisting of regular lattices.
To demonstrate this, the procedure is repeated for rectangular geometry.
The scaling of irregularly shaped crystals is also considered, and again three branches are found ($A^{1/4}, A^{3/4}, A$).
Based on the scaling properties, a quantity is defined which can be used to classify commensurability in infinite as well as finite contacts.
Finally, the consequences for friction experiments on gold nanocrystals on graphite are discussed.


\end{abstract}

\maketitle

\section{Introduction}

Recent years have witnessed a surge of interest in understanding the
microscopic origin of friction as a result of the increased control in surface preparations and the development of local probes like the Atomic Force Microscopes (AFM).
One of the goals of this research is to understand whether extremely 
low friction can be obtained by an appropriate choice of the sliding conditions. 
In particular, commensurability between the sliding lattices is one of the elements that determine friction.
For a purely incommensurate infinite contact, theoretical arguments suggest that static friction should vanish \cite{vanishingstaticfriction,vanishingfriction2}.
This effect has been called superlubricity \cite{Shinjo,fkphononconsoli}.

Commensurability and incommensurability are defined in terms of lattice parameters.
However, specific orientations can also lead to (mis)matches of lattices in contacts.
Very low friction was found in experiments with finite contacts at very low velocities to depend strongly on the orientation~\cite{Dienwiebel2004,torqueandtwist}.
Coexisting states of very different friction have also been observed in the sliding of antimony nanoparticles \cite{Dietzel2008,dietzelprb} and have been attributed to contamination or (in)commensurate interfaces.
Meanwhile, recent theoretical studies~\cite{onsgraphiteflakes,onssquareflakes} have shown that nanocrystals can slide with constant orientation only for particular orientations.

This paper examines systematically theoretically the (in)com\-men\-su\-ra\-bility and friction of sliding nanocrystals with a triangular lattice symmetry [such as Au (111)] on a triangular or hexagonal substrate (such as graphite).
Gold nanocrystals sliding on graphite are a prototype system
for friction
that is being investigated both experimentally~\cite{dietzeltbp} and computationally~\cite{vanossinature2010}.
In this work, the friction is investigated analytically through the total potential energy of the contact layer on the substrate.
The potential energy corrugation plays an essential role in the survival and appearance of stable (in)commensurate sliding orientations~\cite{onsgraphiteflakes}, as well as the order of magnitude of the friction.
In the regime of low sliding velocity, and low temperature, which is the typical situation in AFM experiments~\cite{Schirmeisen2009}, the friction is of the order of $\Delta V \pi/l$, with $l$ the substrate period and $\Delta V$ the corrugation of the total potential energy of the interface.

\begin{figure}
\includegraphics[width=8.6cm]{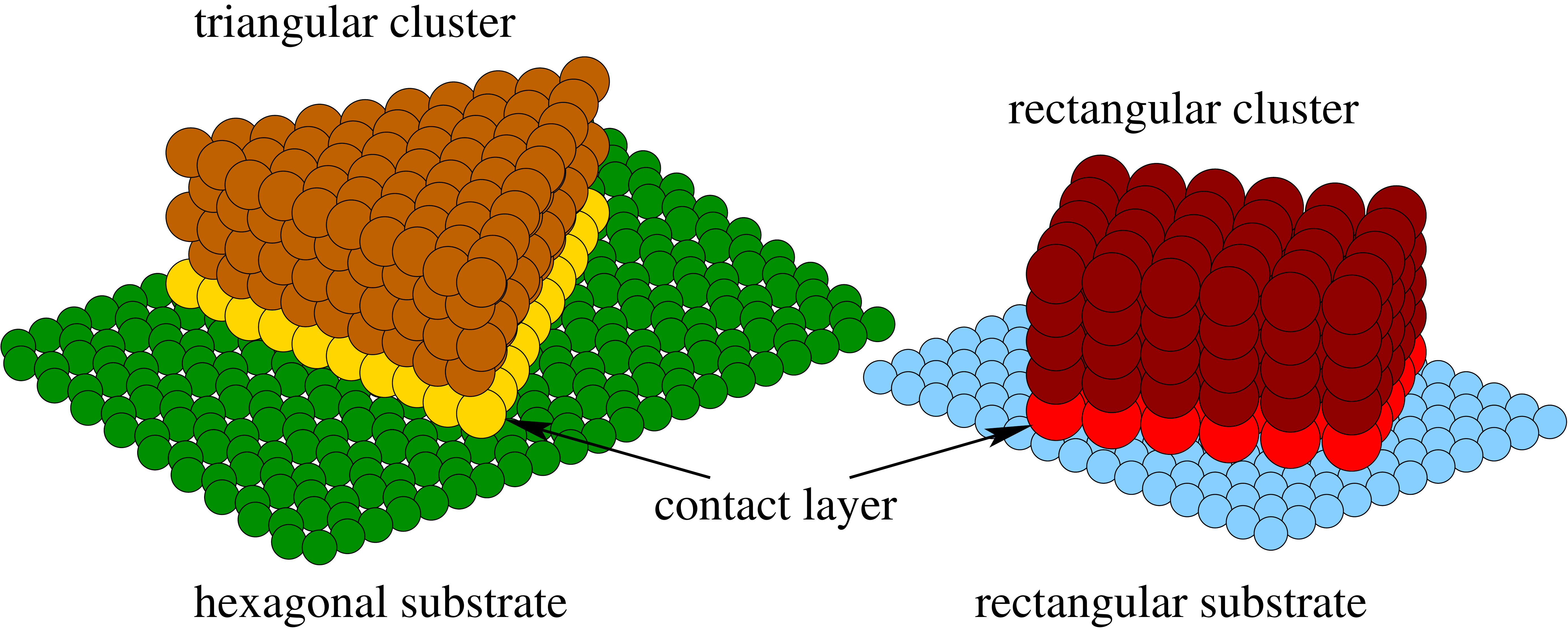}
\caption{
\label{fig:cartoon}
An illustration of the systems under study.  A nanocrystal lies on a triangular or hexagonal substrate with a contact layer that has triangular lattice symmetry.
Systems with rectangular lattice symmetry are discussed as an example in the Appendix.
}
\end{figure}

The interaction between the surface and an atom of the contact layer is modelled with a realistic static potential.
The scaling of the friction with contact size is determined, depending on the orientation, and conclusions are drawn from this regarding commensurate and incommensurate orientations.
The calculations are repeated also for contacts between crystals with rectangular lattice geometries.
An illustration of the systems is shown in Fig.~\ref{fig:cartoon}.

In Sec.~\ref{sec:geometry} the geometry of gold on graphite is introduced.
An expression for the potential energy
is derived in Sec.~\ref{sec:maths}.
At first, rigid crystals are considered, followed by correction terms for crystals with relatively stiff, but deformable latices.
In Sec.~\ref{sec:scaling}, the scaling of the friction is discussed at different (in)commensurate orientations and orientations with different scaling are identified and classified.
It is found that for perfect crystals there are three types of (in)commensurability and consequently three possibilities for the scaling of the friction with the size of the contact area $A$, namely $A^0, A^{1/2}$, and $A$.
Conditions for each are derived.

These three different scaling behaviors for the friction are a direct consequence of the two-dimensional nature of the contact layer, and are thus possible in all geometries, not just triangular.
In the Appendix, the calculations are repeated for a rectangular geometry, to exemplify this.

The scaling of friction for imperfectly and irregularly shaped nanocrystals is discussed in Sec.~\ref{sec:imperfectscaling}.
In Sec.~\ref{sec:discussion}, the implications for experiments are discussed, and a method is proposed for quantifying the commensurability of finite-size crystal interfaces.
Finally, conclusions are presented in Sec.~\ref{sec:conclusions}.

\section{Graphite and gold\label{sec:geometry}}

The interaction between the graphite substrate and the gold crystal are dominated by the bottom, contact layer of the gold crystal and the top layer of the graphite.
Commensurability of these two layers is not only controlled by the ratio of the lattice parameters, but also by their relative orientation.

The surface of graphite has a hexagonal lattice of carbon atoms with an inter-atomic distance of $a=1.42~\mathrm{\AA}$.
A two-dimensional hexagonal potential for a gold atom at postition $\vec{r}$ on a graphite substrate can be written as
\begin{align}
V_\mathrm{Au}(\vec{r}) = - \frac29 V_0\sum_{l=1}^3\cos\left(\frac{2 \pi}{\alpha } \vec{e}_l\cdot\vec{r}\right)~,
\label{eq:substratepotential}
\end{align}
where $\vec{e}_l = \cos[(l-1) \pi/3] \vec{e}_x + \sin[(l-1) \pi/3] \vec{e}_y$ is the unit vector in the direction $(l-1) \pi/3$ with respect to the $x$-axis, and $\alpha = a\frac32$.
Without any load force, the potential corrugation of a single gold atom on graphite is denoted by $V_0$ and is around $50~\mathrm{meV}$~\cite{goldgraphitediffusionbarrier}.
A load force can be accounted for with a higher corrugation.
With a negative $V_0$, this potential can also be used to describe a substrate with a triangular lattice.
While in general a substrate potential may not be sinusoidal, it is always periodic and hence can be written as a sum of sinusoidal potentials.  The analysis described in this work can also be applied to sums of different sinusoidal potentials.
More details of the geometry are shown in Fig.~\ref{fig:trapezoid}.

From the triangular shape of the nanocrystal found in experiments~\cite{dietzeltbp}, it can be deduced that the contact layer must have similar triangular symmetry.
As bulk gold has an fcc lattice, it is reasonable to conclude that the contact layer must be the (111) cleavage.
In the bulk, the lattice parameter is $B'=4.080$~\AA, and consequently the inter-atomic distance in the (111) layer is $B=2.885~\mathrm{\AA}$.
However, the Au(111) surface is known to show surface reconstruction under a wide range of conditions~\cite{aureconstruction}.
In vacuum, the reconstructed Au(111) surface looks similar to the unreconstructed bulk (111) layers, but has a superstructure with a unit cell of size $23B\times\sqrt{3} B$ 
and an average compression in the (110) direction.
Consequently, the rotational symmetry of the contact layer is lost.
In addition, the surface gold atoms do not lie in a plane, but are shifted by up to 0.15~\AA.

As the interaction between the graphite substrate and the Au atoms is weak compared to the interaction between the Au atoms, the reconstruction of the contact layer in the Au-graphite configuration is likely similar to that of Au(111) in vacuum.
In this work, we simply assume that the lattice is only slightly distorted, with the atoms arranged in scalene triangles.
As the interaction between the graphite surface and the gold atoms is weak, the shift of 0.15~\AA\ orthogonal to the surface does not lead to significant changes in the interactions.
As the results derived in this work can be generalized to any regular lattice, they can be applied to any reconstructed surface.

In general, the nanocrystal can be flexible, and may be deformed due to forces exerted by the substrate.
In this work, the crystal is at first assumed to be rigid, and then corrections are made for the displacements of atoms with respect to their equilibrium position.
Elastic deformations can be neglected~\cite{perssontosatti,Sokoloff2000,mueserstructurallubricity}, as the typical coherence length of gold, about 1$\mu$m, is much larger than the size of a crystal in friction experiments~\cite{dietzeltbp}.

Young's modulus, which for gold is $E_\mathrm{Au}=79$~GPa, gives the stress-strain response and can be used as an indication of the displacement $\delta \vec{r}$ of a single atom with respect to the equilibrium lattice positions.
One may consider the force exerted by the substrate on an atom at position $\vec{r}$, $\vec{F}_\mathrm{Au}(\vec{r})$
to work on a single unit cell of the contact layer, of diameter $B$.
The displacement of the atom can be estimated as
\begin{align}
\label{eq:displacementFestimate}
\delta \vec{r} \sim B \frac{\vec{F}_\mathrm{Au}(\vec{r})}{B^2 E_\mathrm{Au}}~.
\end{align}
This leads to a typical displacement of a single gold atom in the crystal of about $0.007$~\AA, corresponding to a difference in potential energy two orders of magnitude smaller than the corrugation.
Though Young's modulus is a bulk property, and we are interested in the displacement of surface atoms, it is sufficient for obtaining the order of magnitude.
A similar estimate for the order of magnitude of the displacement of a surface atom can be obtained from the total number of neighboring atoms and parameters used in molecular-dynamics simulations with atomistic force fields, such as that of Ref.~\cite{goldforcefield}.
In the next section, for the total potential energy of the contact layer, the contribution from the displacement of surface atoms are neglected.
In Sec.~\ref{sec:systematic_soft}, 
corrections that include the displacement of the atoms with respect to their equilibrium positions are also made.

\section{Potential energy landscape\label{sec:maths}}

It has been shown
 that sliding crystals rotate to specific orientations, which are stable and remain (nearly) constant for all time~\cite{onsgraphiteflakes,onssquareflakes}.
These orientations can be identified with periodic orbits in the dynamics and obtained from the potential energy landscape.
They occur when the potential energy averaged over a scan line and its corrugation simultaneously exhibit an extremum as a function of orientation.
However, the incommensurate orientations typically have higher average energy than the commensurate ones, and can be destroyed by sufficiently large thermal fluctuations, leading to commensurate sliding and an increase of friction.
In experiments of small graphite flakes on graphite, incommensurate orientations were found to decay~\cite{torqueandtwist}.
Larger crystals were found theoretically to rotate more slowly, so that incommensurate orientations survive.
The sliding gold nanocrystals studied in the experiments by Dietzel et al.~\cite{dietzeltbp} are sufficiently large for this (with contact areas between $10^3$ and $10^5$ nm$^2$), though the smaller crystals studied numerically by Guerra et al.~\cite{vanossinature2010}, can rotate easily while sliding.
In order to determine the scaling of the friction, and hence the commensurate orientations, as well as apply the theory of Ref.~\cite{onsgraphiteflakes},
it is necessary to focus on the potential energy of the contact layer on the substrate as a function of position and relative orientation.

\begin{figure}
\begin{center}
\includegraphics[width=8.6cm]{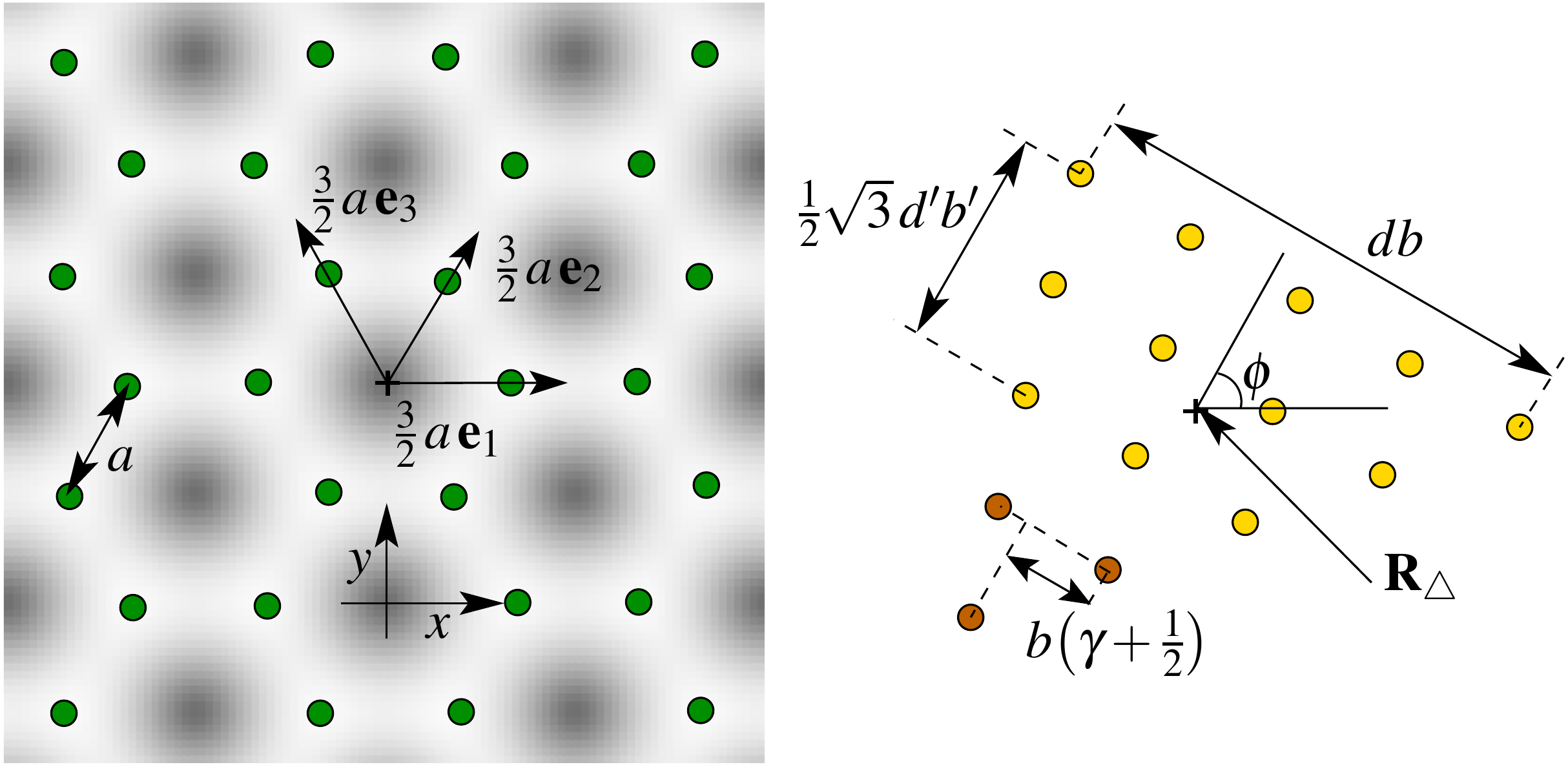}
\end{center}
\caption{
\label{fig:trapezoid}
Diagram of the hexagonal graphite substrate lattice (left) and a trapezoidal gold crystal (right) with definitions of the size parameters $d, d^\prime$, lattice spacings $a, b, b^\prime$, shape parameter $\gamma$, substrate lattice vector $\vec{e}_1$, orientation $\phi$, and position $\vec{R}_\triangle$.
The dark atoms shown complete the trapezoid crystal into a triangular one.
We use $b^\prime/b = \beta$ and $\alpha = \tfrac32 a$.
For gold, with the reconstruction that has been observed in vacuum~\cite{aureconstruction}, the parameters for the lattice are $b=2.760$~\AA, $b^\prime = 2.885$~\AA, $\beta = 22/23$, $\gamma=0$.
In a perfectly triangular crystal, $d=d^\prime, \beta=1,\gamma=0$.
}
\end{figure}

Let us calculate the total potential energy of a trapezoidal crystal.
The geometry of such as crystal is displayed in Fig.~\ref{fig:trapezoid}, along with definitions of the diameters $d$ and $d^\prime$, the inter-atomic distances $b$ and $b^\prime$, the shape parameter $\gamma$, the orientation $\phi$, and the position of the center of smallest triangle that contains the trapezoid, $\vec{R}_\triangle$.
First, we assume that the crystal is rigid, and then make a systematic correction for displacement of atoms from their equilibrium position.

The total potential energy of the crystal at position $\vec{R}$ and orientation $\phi$
on the substrate is dominated by the interaction of the contact-layer atoms with the periodic substrate potential.
It can therefore be written as a sum over all atoms in the contact layer,
\begin{align}
V_{\mathrm{trap.}, d,d^\prime,\alpha,\beta,\gamma}(\vec{R},\phi) = \sum_{j=0}^{d^\prime} \sum_{k=0}^{d-j} V_\mathrm{Au}(\vec{R}_\triangle+\vec{r}_{j,k})~,
\end{align}
where $\vec{r}_{j,k}$ are the positions of the atoms in the contact layer relative to $\vec{R}_\triangle$, the center of the contact layer.

\subsection{\label{sec:rigidpotential}Rigid crystal}

If the crystal is rigid, the positions of the atoms can be written as
\begin{align}
\vec{r}_{j,k} = &b {\mathcal R}(\phi) \cdot \left(\tfrac12 \sqrt{3} \left(\tfrac13 d -j\right),\right.\nonumber\\
&\left.\phantom{\sqrt{3}}\null -\tfrac12 \beta (d-j)+\beta k + \gamma (\tfrac13 d-j) \right)~,
\end{align}
with ${\mathcal R}(\phi)$ the $2\times2$ matrix that performs a rotation over an angle $\phi$.

By substituting the substrate potential, Eq.~(\ref{eq:substratepotential}), and rewriting the $\cos$ in terms of imaginary exponentials, we find that for a rigid crystal,
\begin{widetext}
\begin{align}
V_{\mathrm{trap.}, d,d^\prime,\alpha,\beta,\gamma}^{(0)}(\vec{R},\phi) & =
- \tfrac19 V_0 \sum_{l=1}^3 S_{d,d^\prime,\alpha,\beta,\gamma}(\vec{R}_\triangle,\phi,l)~,\\
S_{d,d^\prime,\alpha,\beta,\gamma}(\vec{R}_\triangle,\phi,l) & = \sum_{j=0}^{d^\prime}   \sum_{k=0}^{d-j} \left[ \exp\left(\frac{2 \pi i }{\alpha } \vec{e}_l\cdot(\vec{R}_\triangle+\vec{r}_{j,k})\right) + \exp\left(-\frac{2 \pi i }{\alpha } \vec{e}_l\cdot(\vec{R}_\triangle+\vec{r}_{j,k})\right)\right]~.
\label{eq:S}
\end{align}
Due to the rotational symmetry of the substrate and of a perfectly triangular crystal,
\begin{align}
S_{d,d^\prime,\alpha,\beta,\gamma}(\vec{R}_\triangle,\phi,l+1)& = S_{d,d^\prime,\alpha,\beta,\gamma}\left({\mathcal R}\left(-\tfrac13 \pi\right)\cdot\vec{R}_\triangle,\phi-\tfrac13 \pi,l\right)~,\\
S_{d,d,\alpha,1,0}(\vec{R}_\triangle,\phi,l+1) &= S_{d,d,\alpha,1,0}\left({\mathcal R}\left(-\tfrac13 \pi\right)\cdot\vec{R}_\triangle,\phi,l\right)~.
\end{align}
Consequently, for an irregular trapezoid, it suffices to determine $S_{d,d^\prime,\alpha,\beta,\gamma}(\vec{R}_\triangle,\phi,1)$.
Finally, the sums in Eq.~(\ref{eq:S}) can be evaluated by making use of the relation
\begin{align}
\sum_{j=0}^{n} \exp\left({i2 c j}\right) = \exp\left({i c n} \right) {\sin\left[{c (n+1)} \right]}/{\sin\left({c} \right)}~,
\label{eq:sinratio}
\end{align}
with $c$ any real number.
One finds that
\begin{align}
S_{d,d^\prime,\alpha,\beta,\gamma}(\vec{R}_\triangle,\phi,1) & = \sum_{j=0,1} \sin\left\{ \frac{\pi b}{\alpha}\left[ \frac{\vec{R}_\triangle\cdot\vec{e}_1}{b} + \tfrac12 (-1)^j \left(2d-d^\prime+2 \right) \beta \sin \phi + \tfrac13 (2d-3d^\prime) \gamma \sin\phi + \tfrac1{6} \sqrt{3} (2d-3d^\prime) \cos\phi\right]\right\}\nonumber\\
&\phantom{\sum_{j=0,1}}\times\frac{\sin\left\{ \frac{\pi b}{\alpha} (d^\prime+1) \left[ \tfrac12 \sqrt{3} \cos\phi + \left(\gamma + \tfrac12 (-1)^j \beta \right) \sin\phi \right]\right\}}{\sin\left\{ \frac{\pi b}{\alpha} \left[ \tfrac12 \sqrt{3} \cos\phi + \left(\gamma + \tfrac12 (-1)^j \beta\right) \sin\phi \right]\right\} \sin\left(\frac{\pi b}{\alpha} (-1)^j \beta \sin\phi\right)}~,
\label{eq:pottrap2}
\\
V_{\mathrm{trap.}, d,d^\prime,\alpha,\beta,\gamma}^{(0)}(\vec{R},\phi)
&=
\null- \tfrac19 V_0 \sum_{l=1}^3 \sin\left[ \frac{\pi b}{\alpha}\left( \frac{\vec{R}_\triangle\cdot\vec{e}_l}{b}\right) \right]
S_{d,d^\prime,\alpha,\beta,\gamma}\left(-\tfrac12 \alpha\vec{e}_l ,\phi,l\right)
\nonumber\\
\label{eq:pottrap1}
&\phantom{= \null}{}\null
- \tfrac19 V_0 \sum_{l=1}^3 \cos\left[ \frac{\pi b}{\alpha}\left( \frac{\vec{R}_\triangle\cdot\vec{e}_l}{b}\right) \right]
S_{d,d^\prime,\alpha,\beta,\gamma}\left(0,\phi,l\right)
~.
\end{align}
Note that the angle-dependent corrugation prefactor  contains all the geometric dependence on size and orientation, and does not depend on the position $\vec{R}$ of the center of mass on the substrate.
The total corrugation is
\begin{align}
\label{eq:pottrapcor}
\max_{\vec{R}}(V^{(0)})- \min_{\vec{R}}(V^{(0)})&= \tfrac13 {V_0}\left[
    S_{d,d^\prime,\alpha,\beta,\gamma} \left( -\tfrac12 \alpha\vec{e}_1 ,\phi,1 \right)^2
    + S_{d,d^\prime,\alpha,\beta,\gamma} \left(0 ,\phi,1\right)^2 \right]^{\tfrac{1}{2}}
 \left( \tfrac32 \cos \frac{\xi}{3} + \tfrac12 \sqrt{3} \sin \frac{|\xi|}{3} \right)~,\\
\xi &= \arctan\left(\frac{S_{d,d^\prime,\alpha,\beta,\gamma} \left( -\tfrac12 \alpha\vec{e}_1 ,\phi,1 \right)}{S_{d,d^\prime,\alpha,\beta,\gamma} \left(0 ,\phi,1\right)^2}\right)~,
\end{align}
\end{widetext}
where $\max_\vec{R}$ and $\min_\vec{R}$ are used to denote the value of the maxima and minima of the potential respectively.

The average potential energy over any scan line vanishes for all scan lines not exactly orthogonal to one of the $\vec{e}_l$.
In this case, the conditions for the existence of stable orientations of Ref.~\cite{onsgraphiteflakes} are met.
As is also the case for rectangular lattices in Ref.~\cite{onssquareflakes}, the incommensurate stable orientations are distributed evenly over the unit circle and their number grows linearly with the diameter~$d$.
It should be noted that these results are consistent with the numerical results for the distribution of activation energies described in Ref.~\cite{Manini2011} for rigid crystals with triangular symmetry on a square lattice.
However, because those results were obtained for relatively small clusters, the authors could not distinguish the commensurate orienations with larger corrugation.

\subsection{Non-rigid crystal\label{sec:firstordercorrections}\label{sec:systematic_soft}}

In realistic nanocrystals, atoms of the contact layer are displaced with respect to their equilibrium position, due to the forces exerted on them by the substrate.
From the arguments presented in Sec.~\ref{sec:geometry}, it is clear that this displacement is small for the system studied in this work.
However, as the contact layer of typical nanocrystals contain many atoms, the energy effect is not necessesarily negligeable compared to the low total potential energy for incommensurate configurations.

To provide some insight into the effects of lattice deformations without making the calculations much more complicated, it is assumed here that each atom couples only to its equilibrium position.
Within this mean-field approximation, small displacements $\delta\vec{r}$ from the equilibrium position are due to a force equal to
\begin{align}
\vec{F}_\mathrm{spring} = \kappa \delta\vec{r}~,
\end{align}
where the spring constant $\kappa$ can be estimated from Eq.~(\ref{eq:displacementFestimate}) to be,
\begin{align}
\kappa \sim B E_\mathrm{Au} = 1.4 \times 10^3~\mathrm{meV/\AA^2}~.
\end{align}

Because typical sliding velocities in experiments are low, we may limit ourselves to the quasi-static case, where all forces on each atom sum up to zero.
The correction to the potential energy consists of the reduction in potential energy due to the displacement on the lattice, and the additional potential energy stored in the spring.
To leading order in the displacement, we find a correction to the potential energy of the atom with equilibrium position~$\vec{r}$~of
\begin{align}
\Delta V^{(1)}_\mathrm{Au}(\vec{r}) = - \frac1{2\kappa} |\vec{F}_\mathrm{Au}(\vec{r})|^2~,
\end{align}
with $\vec{F}_\mathrm{Au}(\vec{r})= - \partial V_\mathrm{Au}(\vec{r})/\partial \vec{r}$ the force exerted by the substrate on an atom at position $\vec{r}$.

By substituting Eq.~(\ref{eq:substratepotential}), using the relation $2 \sin {\mathcal X} \sin {\mathcal Y} = \cos({\mathcal X}-{\mathcal Y})-\cos({\mathcal X}+{\mathcal Y})$, and using the properties of $\vec{e}_l$, one obtains
\begin{widetext}
\begin{align}
\Delta V_\mathrm{Au}(\vec{r}) & = -\frac1{2\kappa} \left[ \frac{2}{9} V_0 \sum_{l=1}^3 \frac{2 \pi}{\alpha } \sin \left(\frac{2 \pi}{\alpha } \vec{e}_l\cdot\vec{r}\right) \vec{e}_l\right]^2\\
& = - \frac{4 \pi^2 V_0^2}{81\kappa \alpha^2} \sum_{l=1}^3 \left[
1 - \cos\left(\frac{4 \pi}{\alpha } \vec{e}_l\cdot\vec{r}\right)
+\cos\left(\frac{2 \pi}{\alpha } \vec{e}_{l} \cdot\vec{r}\right)
-\cos\left(\frac{2 \sqrt{3}\,\pi}{\alpha} \vec{e}_{l+\tfrac12} \cdot\vec{r}\right)
\right]
\end{align}
This expression is of a form very similar to Eq.~(\ref{eq:substratepotential}).
The approach used in Sec.~\ref{sec:rigidpotential} to obtain the total potential energy of a rigid crystal can be applied to this expression as well.
Hence, one obtains the first-order correction of
\begin{align}
\Delta V^{(1)}_{\mathrm{trap.}, d, d^\prime,\alpha,\beta,\gamma}(\vec{R},\phi) &= \frac{\pi^2 V_0}{9\kappa \alpha^2}
\left[
\tfrac{2}{9} V_0
+ V_{\mathrm{trap.}, d,d^\prime,\alpha,\beta,\gamma}^{(0)}(\vec{R},\phi)
\right.\nonumber\\ 
& \phantom{= \frac{2 \pi^2 V_0}{9\kappa \alpha^2} []} \left.
\null
- V_{\mathrm{trap.}, d,d^\prime,\alpha/2,\beta,\gamma}^{(0)}(\vec{R},\phi)
- V_{\mathrm{trap.}, d,d^\prime,\alpha/\sqrt{3},\beta,\gamma}^{(0)}({\mathcal R}(\pi/6) \cdot \vec{R},\phi+\pi/6)
\right]
~.
\label{eq:firstordercorrection}
\end{align}
\end{widetext}
Though the functional forms of the terms in this correction are very similar to Eq.~(\ref{eq:pottrap1}), the values of the parameters are different.
As is discussed in the next section, this allows the first-order correction to become larger than the leading order in some cases, despite the small prefactor.

Higher order corrections and corrections for direct coupling to the neighboring atoms in the contact layer
could, in principle, be obtained using the same approach.
However, these corrections would have to include the nonlinear response of the lattice, as well as second-order derivatives of the potential energy.
This would lead to expressions similar in form to Eq.~(\ref{eq:substratepotential}), and therefore to additional terms similar to Eq.~(\ref{eq:firstordercorrection}), but with yet again different parameters and smaller prefacors.

\section{Scaling and orientation\label{sec:scaling}}

From the structure of the function $S$ in Eq.~(\ref{eq:pottrap2}), one can derive important general geometrical and scaling properties.
From Eqs.~(\ref{eq:pottrap2}),~(\ref{eq:pottrap1}), and~(\ref{eq:firstordercorrection}) it is clear that there are a number of parameters which control the potential energy corrugation of the surface which the nanocrystal is subjected to.
For a perfectly triangular crystal these are the diameters $d$, the ratio of the lattice constants, $b/\alpha$, and the orientation $\phi$.

\subsection{Diverging denominator}

\begin{figure}
\includegraphics[angle=270,width=8.6cm]{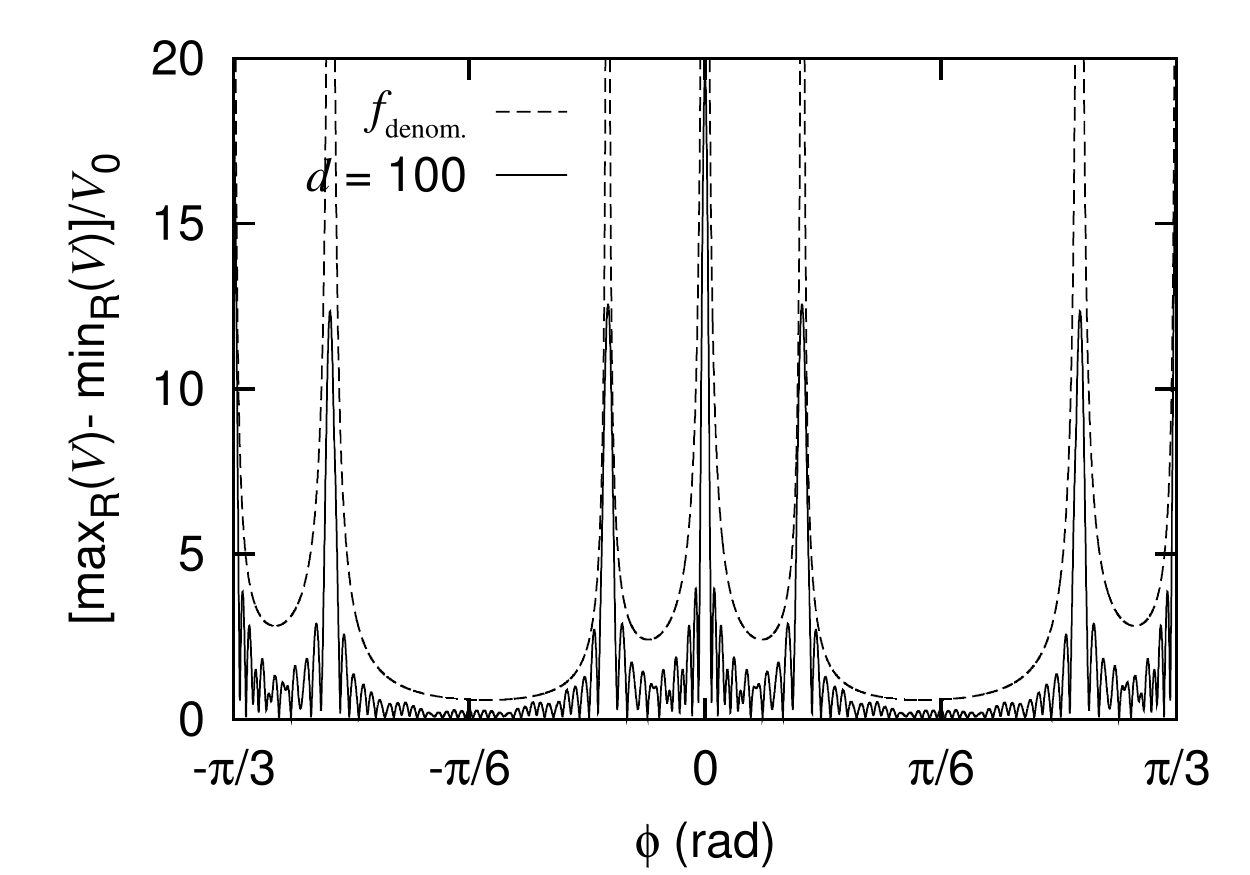}
\caption{
\label{fig:numdenom}
The angle-dependent corrugation prefactor, Eq.~(\ref{eq:pottrapcor}), for a perfectly triangular nanocrystals on a hexagonal substrate with $d=100, \alpha = 1.354$ (solid line), along with $f_{\mathrm{denom.},\alpha}$ (dotted line), which does not depend on the diameter $d$.
The ratio $b/\alpha\approx 1.325$.
The denominators control the typical size of the corrugation.
Whenever the denominator nearly diverges near the orientations given by Eqs.~(\ref{eq:denominator1}) and~(\ref{eq:denominator2}), a peak appears.  There, the interface is (partially) commensurate.
}
\end{figure}

\begin{figure}
\includegraphics[angle=270,width=8.6cm]{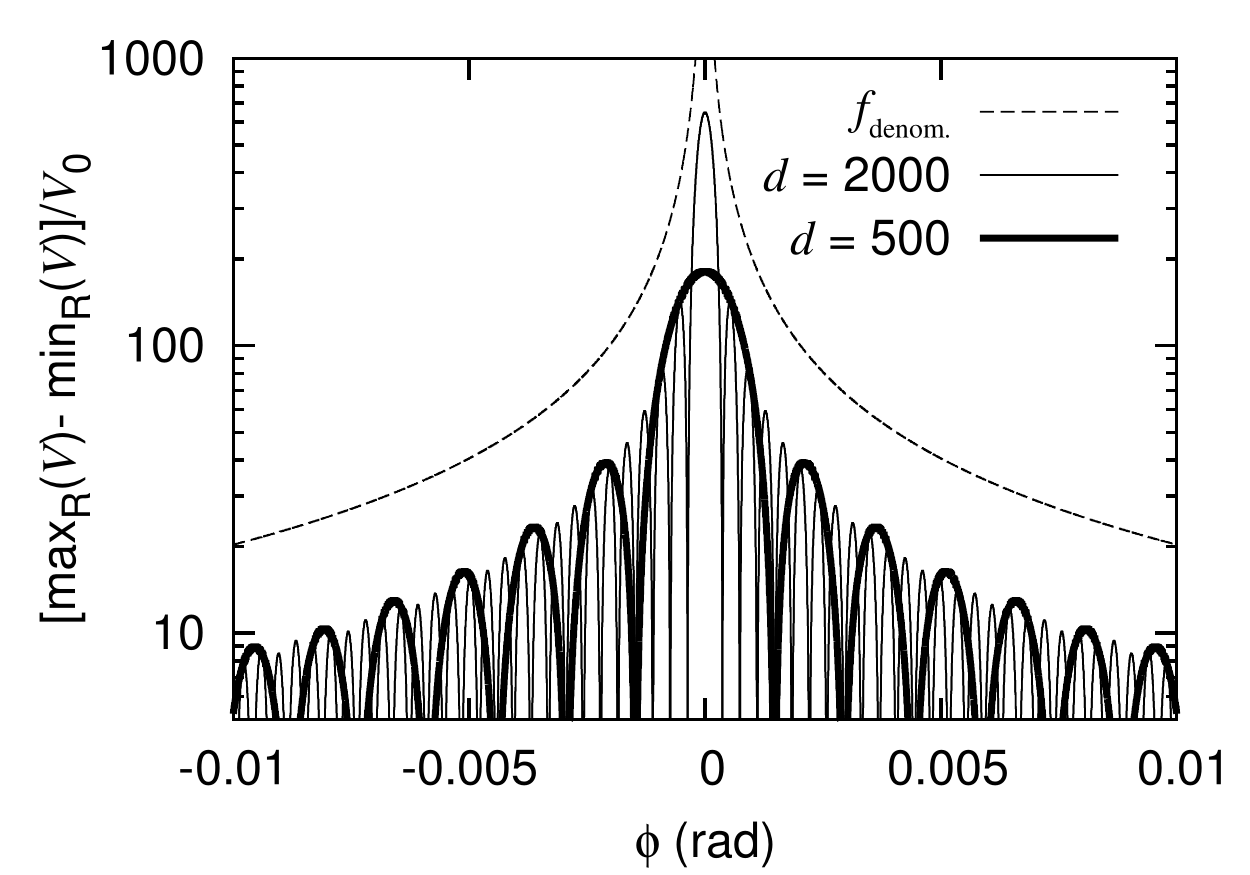}
\caption{
\label{fig:numdenom2}
A logarithmic plot of a peak in the corrugation for a perfectly triangular rigid nanocrystals on a hexagonal substrate with $d=2000, \alpha = 1.354$ (thin solid line) and $d=500$ (thick solid line), along with $f_{\mathrm{denom.},\alpha}$ (dotted line).
The denominators control the typical size of the corrugation.
In this case, the corrugation diverges with $1/(\phi-\phi_\mathrm{c.})$, but, very close to $\phi_\mathrm{c.}$, it remains finite, because the numerator vanishes as well.
}
\end{figure}

In Eq.~(\ref{eq:pottrap2}), the numerator varies rapidly between~-1 and~1, with $\phi$, while the denominator provides a size-indepent prefactor that can diverge for particular angles.
This can be seen in Figs.~\ref{fig:numdenom} and~\ref{fig:numdenom2}, where one of the angle-dependent corrugation prefactors is plotted along with the function
\begin{widetext}
\begin{align}
f_{\mathrm{denom.},\alpha,\beta,\gamma}(\phi)& = \tfrac{1}{9} V_0\left|\sin\left\{ \frac{\pi b}{\alpha} \left[ \tfrac12 \sqrt{3} \cos\phi +\left(\gamma + \tfrac12\right) \beta \sin\phi \right]\right\}
\sin\left(\frac{\pi b\beta}{\alpha} \sin\phi\right)\right|^{-1}
\nonumber\\
& \phantom{=}
\null + \tfrac{1}{9} V_0 \left|\sin\left\{ \frac{\pi b}{\alpha} \left[ \tfrac12 \sqrt{3} \cos\phi +\left(\gamma- \tfrac12 \beta\right) \sin\phi \right]\right\}
\sin\left(- \frac{\pi b \beta}{\alpha} \sin\phi\right)\right|^{-1}~.
\end{align}
\end{widetext}
This function gives an upper bound for the absolute value of the corrugation due to a particular term.
It can be used to estimate the typical corrugation of any triangular or trapezoidal crystal.
At particular combinations of orientations $\phi_\mathrm{c.}$ and lattice parameter ratio $b/\alpha$, one or both of the denominators vanishes and $f_{\mathrm{denom.},\alpha}(\phi)$ diverges.
The numerator in Eq.~(\ref{eq:pottrap2}) vanishes as well, so that $S$ itsself does not diverge.
As the arguments of the trigonometric functions in the numerator contain factors of $d$ and $d^\prime$, the resulting contribution to $S$ is an order of $d$ or $d^\prime$ higher than at other orientations.

There is some arbitraryness to the choice of numerator and denominator, as both can be multiplied by any function.
This cannot however lead to additional factors of $d$.
For determining the order of magnitude of the corrugation at any orientation, $f_{\mathrm{denom.},\alpha}(\phi)$ therefore suffices.

\subsection{Classification of commensurability}

The (in)commensurate configurations can be distinguished and classified by the exponent $C$ with which the corrugation scales with the contact diameter.
This order of the commensurability can be obtained from the divergence of the size-independent prefactor in the corrugation, which goes as $(\phi-\phi_\mathrm{c.})^{-C}$.
I.e., it is equal to the number of derivatives (starting with the zeroth) with respect to $\phi$ of the denominators in Eq.~(\ref{eq:pottrap2}) that are equal to zero.
The scaling of the friction with the contact area can be directly related to the order of commensurability through
\begin{align}
F_\mathrm{fric.}
\propto d^C
\propto A^{\frac{C}{2}}~.
\label{eq:scaling}
\end{align}
For the commensurate orientations generated by the first-order correction in Eq.~(\ref{eq:firstordercorrection}) that do not coincide with commensurate orientations of the leading order, the friction is a factor of order $V_0/(ca)$ smaller.

For most parameter combinations and orientations the contact is incommensurate with $C=0$, the denominator does not vanish at all, and both denominator and numerator are of order $1$.
The corrugation of the nanocrystal on the substrate does not scale with the size of the crystal.
The sliding friction consequently is of the order $V_0/a\propto A^0$.
It should be noted, however, that without a load force, $V_0$ for gold on graphite is comparable to $k_\mathrm{B}T$ at room temperature.
Temperature effects on the friction can therefore play a role as well in this regime, reducing the friction further.

\begin{figure}
\includegraphics[angle=270,width=8.6cm]{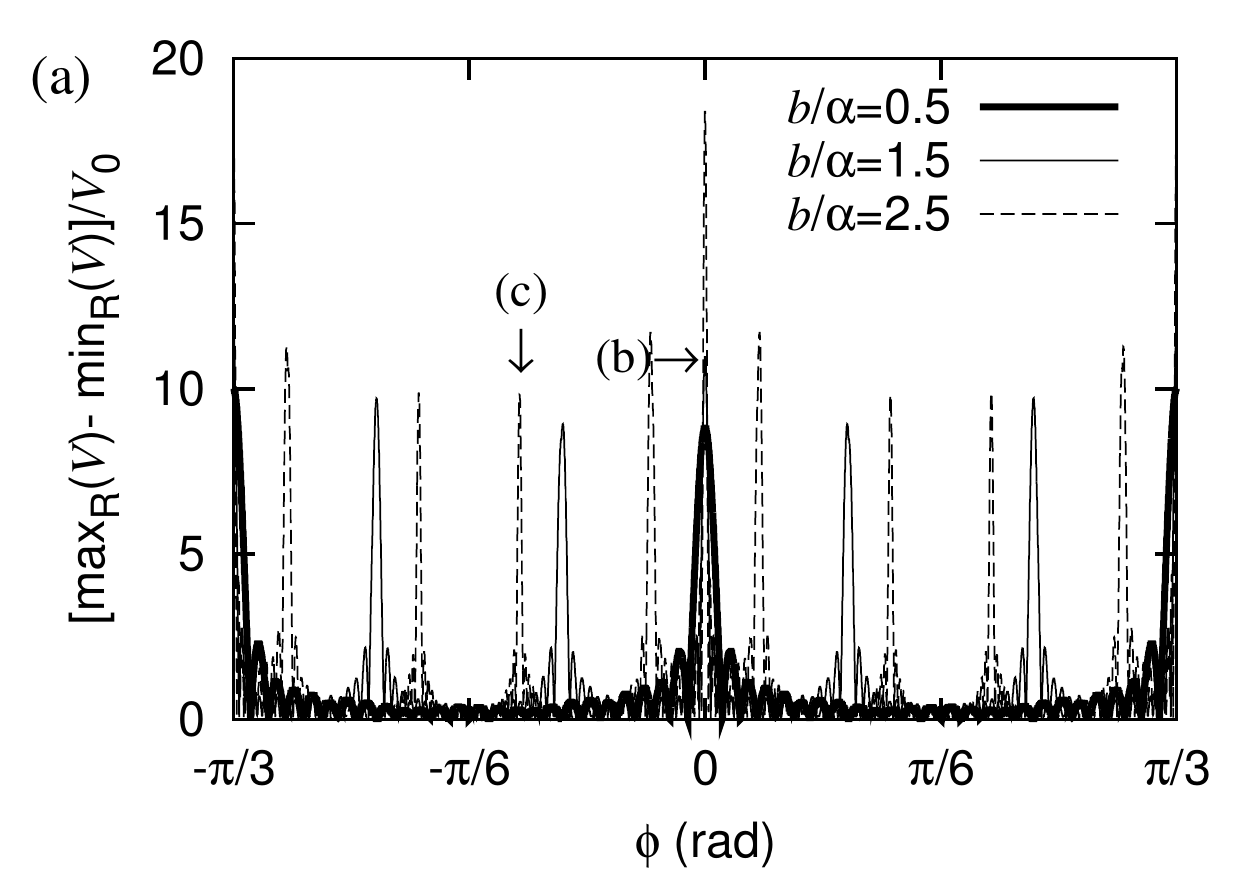}
\vskip\medskipamount
~~\includegraphics[width=8.2cm]{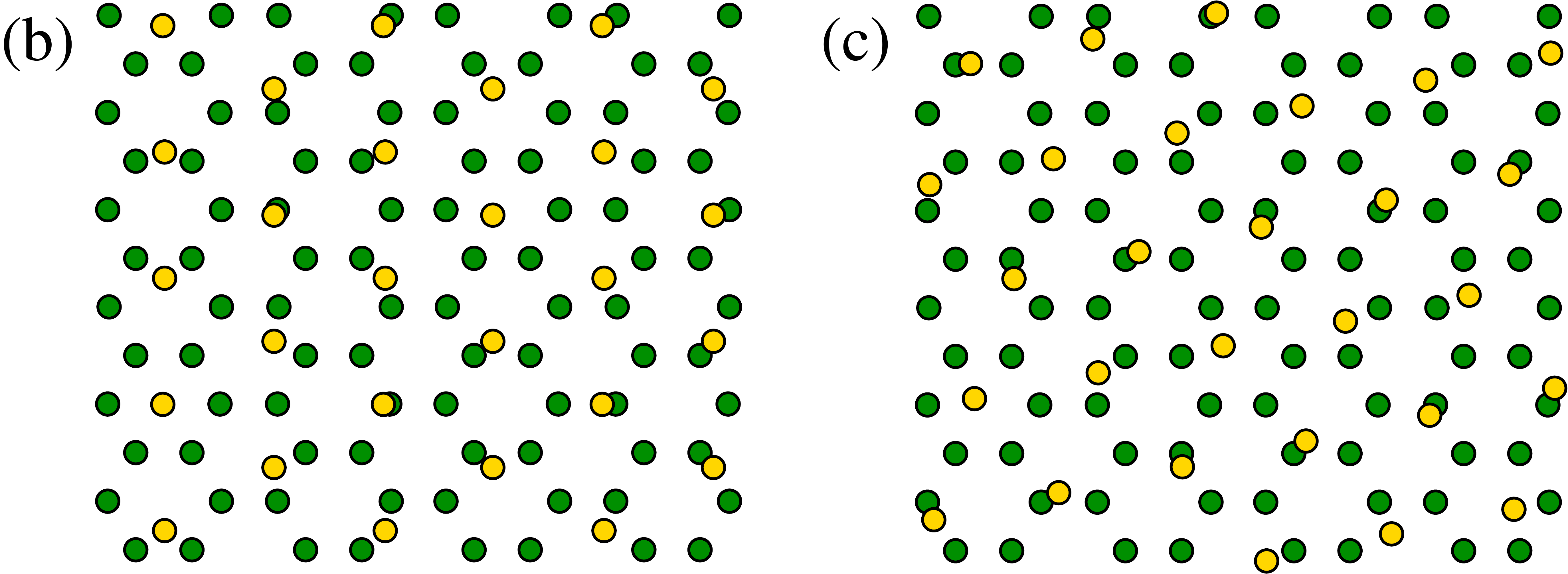}
\caption{
\label{fig:nontrivialangles}
The corrugation (a) as a function of the orientation for crystals with diameter $d=100$ and various $b/\alpha$ and
two commensurate orientations with $C=1$ for triangular crystals with $b/\alpha=1.5$ and with (b) trivial $\phi=0$ and (c) nontrivial $\phi = 0.3175$.
Though the nontrivial commensurate configuration shown in (b) does not have any obvious lining up of symmetry axes, it nevertheless corresponds to a strong peak in the corrugation in (c).
For larger $b/\alpha$, more nontrivial solutions of Eqs.~(\ref{eq:denominator1}) and~(\ref{eq:denominator2}) exist and hence more orientations with large corrugations.
}
\end{figure}

Commensurability order $C=1$ can occur when one of the two $\sin$ functions in the denominator in Eq.~(\ref{eq:pottrap2}) vanishes for a particular orientation $\phi_\mathrm{c.}$.
For a rigid crystal, this occurs when one of the following conditions is met:
\begin{align}
\label{eq:denominator1}
\frac{b}{\alpha} \left[ \tfrac12 \sqrt{3} \cos\phi_{\mathrm{c.}} + \gamma \sin\phi_{\mathrm{c.}} \pm \tfrac12 \beta \sin\phi_{\mathrm{c.}} \right] & = p~,\\
\label{eq:denominator2}
\frac{b\beta}{\alpha} \sin\phi_{\mathrm{c.}} & =p'~,
\end{align}
with $p,p'\in\mathbb{Z}$.
If $b/\alpha$ is sufficiently large, i.~e. large inter-atomic distance in the nanocrystal, solutions exist also for $p\neq 0$.
This implies that there exist nontrivial commensurate-like orientations with high corrugation.
At these orientations, there is no obvious lining up of symmetry axes, or other source of commensurability, as there is for the trivial solutions with $p=0$, but the corrugation and friction nevertheless behave in a similar way.
This is illustrated in Fig.~\ref{fig:nontrivialangles}, where the corrugation is plotted for several values of $b/\alpha$.
Two commensurate combinations of lattices are also shown, one trivial, and one nontrivial.
In the nontrivial one, no lining up is immediately visible, yet the potential energy still scales with the diameter, as it does for the trivial commensurate configuration.
For bulk Au and graphite, $\gamma=0$, $\beta=1$, $b/\alpha\approx 1.354$, and thus there are commensurate orientations $\phi_{\mathrm{c.}} \approx \tfrac13 q \pi \pm 0.216$.

The first-order correction for lattice deformations [Eq.~(\ref{eq:firstordercorrection})]
produces additional commensurate orientations, which satisfy one of the conditions
\begin{align}
\label{eq:denominator3}
\frac{b}{\alpha} \left[ \tfrac12 \sqrt{3} \cos\phi_{\mathrm{c.}} + \gamma \sin\phi_{\mathrm{c.}} \pm \tfrac12 \beta \sin\phi_{\mathrm{c.}} \right] & = \tfrac12 p~,\\
\label{eq:denominator4}
\frac{b\beta}{\alpha} \sin\phi_{\mathrm{c.}} & =\tfrac12 p'~,\\
\frac{b}{\alpha} \left[ \tfrac12 \sqrt{3} \cos(\phi_{\mathrm{c.}}-\tfrac16 \pi)
+ \gamma \sin(\phi_{\mathrm{c.}}-\tfrac16 \pi)
\right.&\nonumber\\\left.\null
\label{eq:denominator5}
\pm \tfrac12 \beta \sin(\phi_{\mathrm{c.}}-\tfrac16 \pi) \right] & = \tfrac13\sqrt{3} p~,\\
\frac{b\beta}{\alpha} \sin(\phi_{\mathrm{c.}}-\tfrac16\pi) & =\tfrac13\sqrt{3} p'~.
\label{eq:denominator6}
\end{align}
These conditions contain additional factors of $1/2$ and $1/\sqrt{3}$ when compared to Eqs.~(\ref{eq:denominator1}) and~(\ref{eq:denominator2}), due to the appearance of such factors in Eq.~(\ref{eq:firstordercorrection}).

\begin{figure}
\includegraphics[angle=270,width=8.6cm]{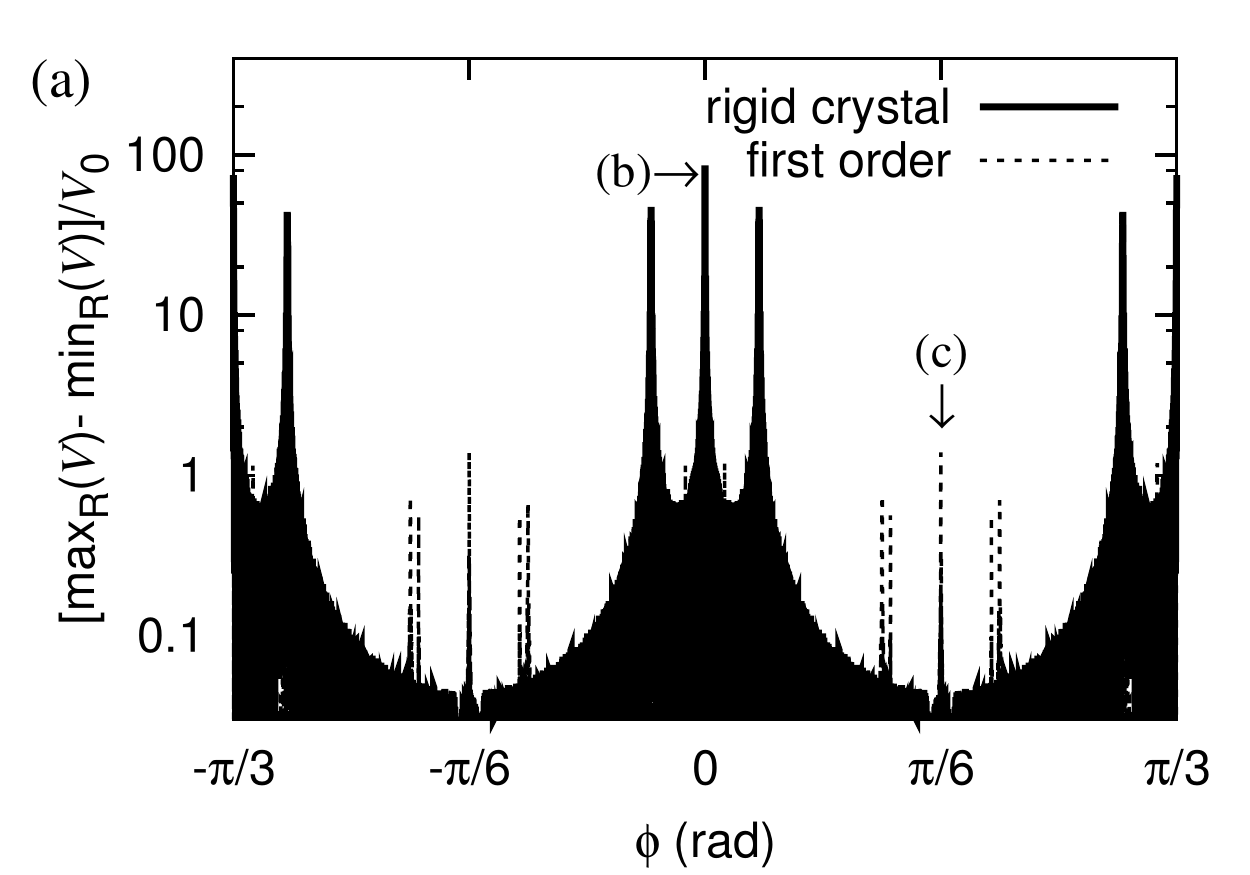}
\vskip\medskipamount
~~\includegraphics[width=8.2cm]{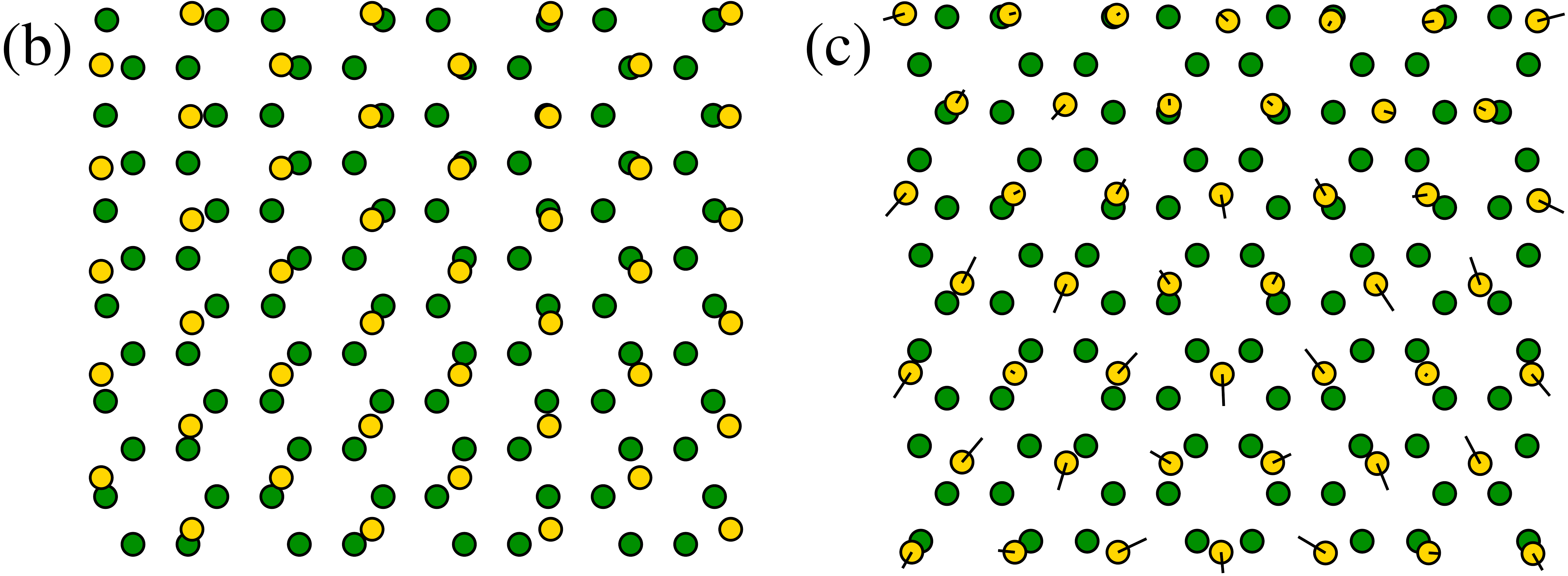}
\caption{
\label{fig:corrections}
The the corrugation (a) as a function of the orientation for non-rigid triangular crystals with diameter $d=1000$ and $2V_0/(9 \kappa \alpha^2)=0.003$ (the typical value for Au),
for triangular crystals with $b/\alpha=5/4$, and
lattices for
commensurate orientations with $C=1$, for (b) a rigid crystal $\phi=0$ and (c) a non-rigid crystal at $\phi = \pi/6$.
The ratio of lattices parameters $b/\alpha=5/4$ is incommensurate.
The short lines in (b) indicate the direction of the displacement of the atoms as a result of the substrate potential, and their length is proportional to its magnitude.
The peaks marked ``rigid crystal'' appear already when the crystal is assumed to be rigid, and satisfy Eqs.~(\ref{eq:denominator1}) and~(\ref{eq:denominator2}).
The lower peaks that appear due to lattice deformations are marked with ``first order'' and are the solutions of Eqs.~(\ref{eq:denominator3}) -- (\ref{eq:denominator6}).
}
\end{figure}

Equations~(\ref{eq:denominator3}) through~(\ref{eq:denominator6}) yield a set of additional commensurate orientations that are different from the ones produced by a rigid crystal.
In Fig.~\ref{fig:corrections}, the corrugation in the total potential energy, consisting of the rigid-crystal term and the first-order correction for lattice deformations, is shown for a very large perfectly triangular crystal.
Peaks corresponding to both the nontrivial commensurate configurations and the additional commensurate configurations associated with lattice deformations are clearly visible.

Finally, $C=2$ can occur only if the two $\sin$ functions in the denominators have coinciding zeros.
This happens only for special ratios of the lattice parameters and corresponds to commensurability in two parameters.
The conditions of Eqs.~(\ref{eq:denominator1}) and~(\ref{eq:denominator2}) for the orientation can only be satisfied simultaneously if the lattice parameters are related by
\begin{align}
\left(\frac{b}{\alpha}\right)^2 = \left(\frac{p'}{\beta}\right)^2 + \frac43 \left[p-p'\left(\frac\gamma\beta \pm \frac12 \right)\right]^2~,
\label{eq:denominatorx1}
\end{align}
with $p,p'\in\mathbb{Z}$.
From the first-order correction in Eq.~(\ref{eq:firstordercorrection}) additional $C=2$ commensurate combinations of parameters are found,
\begin{align}
\label{eq:denominatorx2}
4 \left(\frac{b}{\alpha}\right)^2 = \left(\frac{p'}{\beta}\right)^2 + \frac43 \left[p-p'\left(\frac\gamma\beta \pm \frac12 \right)\right]^2~,\\
3 \left(\frac{b}{\alpha}\right)^2 = \left(\frac{p'}{\beta}\right)^2 + \frac43 \left[p-p'\left(\frac\gamma\beta \pm \frac12 \right)\right]^2~.
\label{eq:denominatorx3}
\end{align}
Equation~(\ref{eq:denominatorx2}) corresponds to commensurate configurations at angles satisfying Eqs.~(\ref{eq:denominator3}) and~(\ref{eq:denominator4}), while Eq.~(\ref{eq:denominatorx3}) corresponds to commensurate configurations at angles satisfying Eqs.~(\ref{eq:denominator5}) and~(\ref{eq:denominator6}).

A similar structure appears also in the expressions for rectangular nanocrystals on rectangular lattices shown in the Appendix and also treated in Ref.~\cite{onssquareflakes}.
This is due to the fact that any sum over atoms arranged in a regular way in nanocrystal, subjected to a periodic potential, can be written in terms of double sums of the form of Eq.~(\ref{eq:sinratio}).
Hence, two $\sin$ functions always appear in the denominator, and similar conditions exist for which the denominator vanishes.
The system studied in Ref.~\cite{onssquareflakes}, W on NaF, however does not have a sufficiently high ratio of lattice parameters to produce nontrivial commensurate orientations.
Consequently, no nontrivial commensurate orientations were observed there.

\subsection{Preferred sliding directions}
In experiments of friction of nanocrystals, the crystals are either pulled along a particular direction~\cite{Dienwiebel2004}, or pushed~\cite{oldantimony,dietzeltbp}.
When crystals are pulled, the direction of their motion is enforced externally.
When the crystals are pushed, they may move in a direction not precisely along to the force, following some preferred sliding direction.
A comparison between Eqs.~(\ref{eq:pottrap1}) and~(\ref{eq:substratepotential}) shows that the potential energy of a nanocrystal depends on the position in a similar way to that of a single atom.
Consequently, the preferred sliding directions are the same as for a single atom, along vectors othogonal to $\vec{e}_l$.
The nontrivial commensurate orienations do not lead to any nontrivial preferred sliding directions.
In addition, the activation energy is always equal to the same fraction of the corrugation.
However, the first-order corrections produce a potential form which is rotated by $\pi/6$.
For some parameter combinations, this may lead to additional preferred sliding directions parallel to $\vec{e}_l$.

\subsection{Less rigid crystals and the Frenkel-Kontorova Model}
It is interesting to compare the model used here and the results to what is known about the Frenkel-Kontorova (FK) model~\cite{frenkelkontorova} in one and two dimensions.
The FK model consists of a chain or sheet of particles, coupled to one-another and a (quasi-)periodic substrate.
The coupling between the particles may be weak or strong compared to the substrate, leading to different types of behavior.
When the substrate is weak by comparison, for incommensurate lattice parameters, the interface can slide without friction.
In the one-dimensional FK model, commensurate configurations occur for rational ratios of the lattice parameters~\cite{vanishingstaticfriction}.
For the nearly rigid case, in the FK model, there is a cascade of commensurate configurations with decreasing friction.

For weak coupling between monomers in the chain, there is a breaking of analyticity~\cite{vanishingstaticfriction}, which results in any ratio of lattice parameters yielding nonvanishing friction for infinite contacts.
Other effects also start playing a role that do not affect stiff crystals, for instance nucleation of incommensurate domains within a large sheet~\cite{marcoPNAS,marco}).

In the system considered here a similar cascade would appear, if more higher order corrections for lattice deformations were included.
More commensurate orientations with increasingly small friction would be found.
Each subsequent term in the potential energy would however be multiplied by a small prefactor, and so each additional commensurate orientation would correspond to an increasingly small corrugation.

The one-dimensional FK model has only two length scales, and therefore only one parameter which controls commensurability.
The system studied in this work has two parameters which control commensurability, similar to the Frenkel-Kontorova (FK) model in two dimensions, and the quasi-periodic one-dimensional FK model~\cite{quasiperiodicFK}.
Unlike the model described here, the quasiperiodic FK model cannot have coexistence of commensurate states with different order of commensurability.
The two-dimensional FK model, however, is already too complex to be analytically tractable.
Hence, most results dealing with extended two-dimensional deformable interfaces are numerical (see, for instance, Refs~\cite{2dFK-1,2dFK-2,2dFK-3}).
Recently, though, experiments were performed that model atoms as colloids in suspensions, and reproduce the FK model closely~\cite{colloidFK}.

\section{Scaling for imperfectly shaped crystals\label{sec:imperfectscaling}}

In real-world, experimental conditions, crystaline particles are almost never perfectly triangular, or even trapezoidal.
At the very least, even if the crystal lattice is in tact, some atoms may be missing and the corners may be rounded.

Any shape of particle can be written as a sum (including negative terms, if necessary) of regularly shaped particles.
The number of terms in this sum can be estimated from the number of step edges $M$ on the circumference of the contact layer.
Each term contributes an amount to the total potential energy of the order of $V_0 d^C$.
Because the total potential energy at a particular orientation varies rapidly with the size of the particle [see Eq.~(\ref{eq:pottrap2})], one may assume a random phase for
every term.
The total corrugation for an irregular particle is therefore of the order of
$V_0 d^C \sqrt{M}$,
but can never grow faster than $\propto d^2$.
If the shape of the particle is simply scaled, then the number of step edges increases with $M\propto d$.
Such crystals therefore have friction scaling similarly to Eq.~(\ref{eq:scaling}) with,
\begin{align}
F_\mathrm{fric,} \sim \frac{4 \pi}{3 a} V_0 d^{\min(C+1/2,2)} \propto  A^{\min((2C+1)/4,1)}~,
\label{eq:scalingimperfect}
\end{align}
i.~e.\ with exponents $1/4, 3/4$, and $1$.

The effects of lattice distortions at the edge of the contact layer can be included by summing up in a similar way over a small number of contact layers with different sizes to account for the different interactions at the edge of the crystal.
Therefore, the scaling with the crystal size and the (in)commensurate configurations is not affected by the edges.
It should also be noted that, regardless of their shape, the friction of amorphous crystals always scales $\propto d \propto A^{1/2}$~\cite{mueserprl}.

\section{Discussion\label{sec:discussion}}

\subsection{Three friction branches}
\begin{figure}
\includegraphics[angle=270,width=8.6cm]{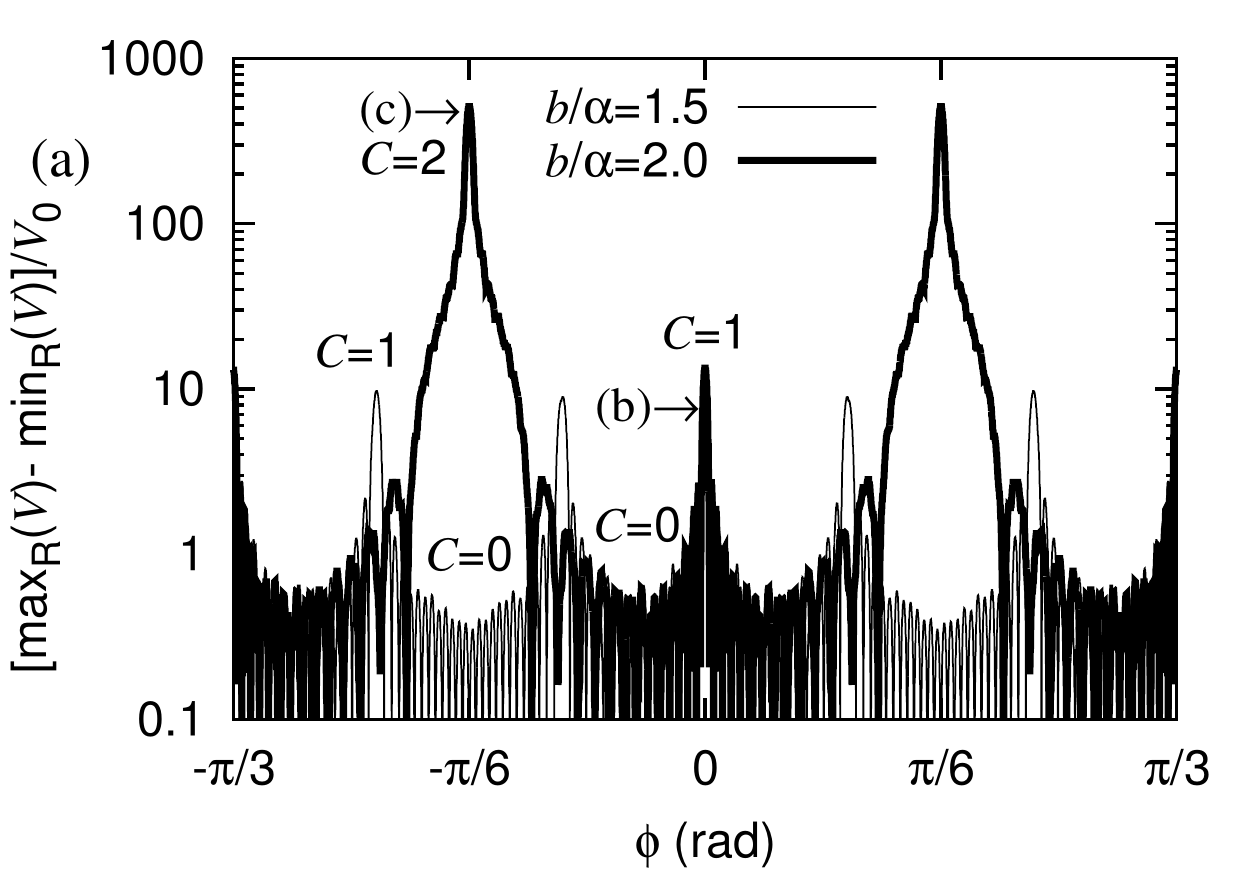}
\vskip\medskipamount
~~\includegraphics[width=8.2cm]{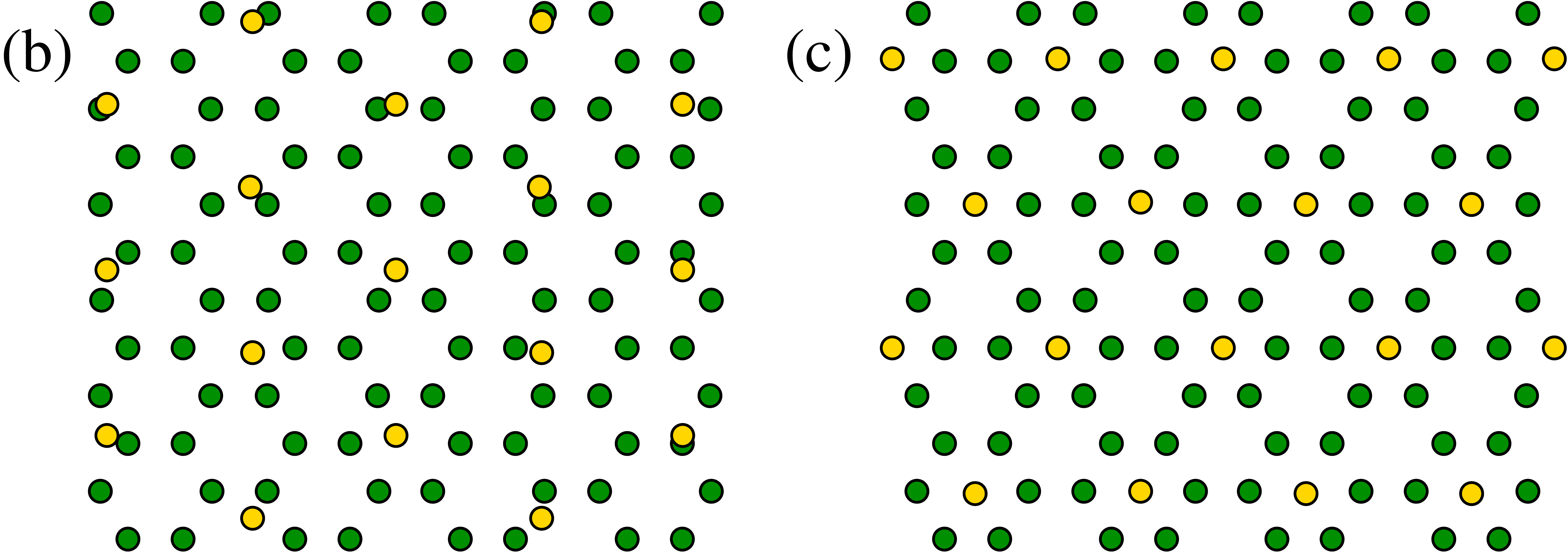}
\caption{
\label{fig:multiplicity}
The corrugation (a) as a function of the orientation for $b/\alpha=2$ and $b/\alpha=1.5$ for a perfectly triangular nanocrystal with $d=50$, and two commensurate orientations for $b/\alpha=2$, at (b) $\phi=0,C=1$ and (c) $\phi=\pi/6,C=2$.
The three different friction branches occur in one interface here.  For the case of $\phi=\pi/6$, the corrugation is of the order $d^2$, while for $\phi=0$ it is of order $d$, and at most other orientations the typical corrugation is not dependent on~$d$.
}
\end{figure}

The three different types of (in)commensurability, quantified with the values of $C=0,1,2$, can occur at different orientations in a single interface already for rigid crystals.
In Fig.~\ref{fig:multiplicity}, such a case is shown.
The trivial commensurate orientations have $C=1$, corrgation scaling with $d$, and are the result of a single vanishing denominator in Eq.~(\ref{eq:pottrap2}).
The nontrivial commensurate orientations are the result of both denominators vanishing at the same orientation, and thus correspond $C=2$ and corrugation scaling with $d^2$.
The corresponding orientations around $\pi/3 q + \pi/6$ are particularly interesting, as these are usually assumed to be strongly incommensurate, and are indeed so for most combinations of lattice parameters.

\begin{figure}
\includegraphics[angle=270,width=8.6cm]{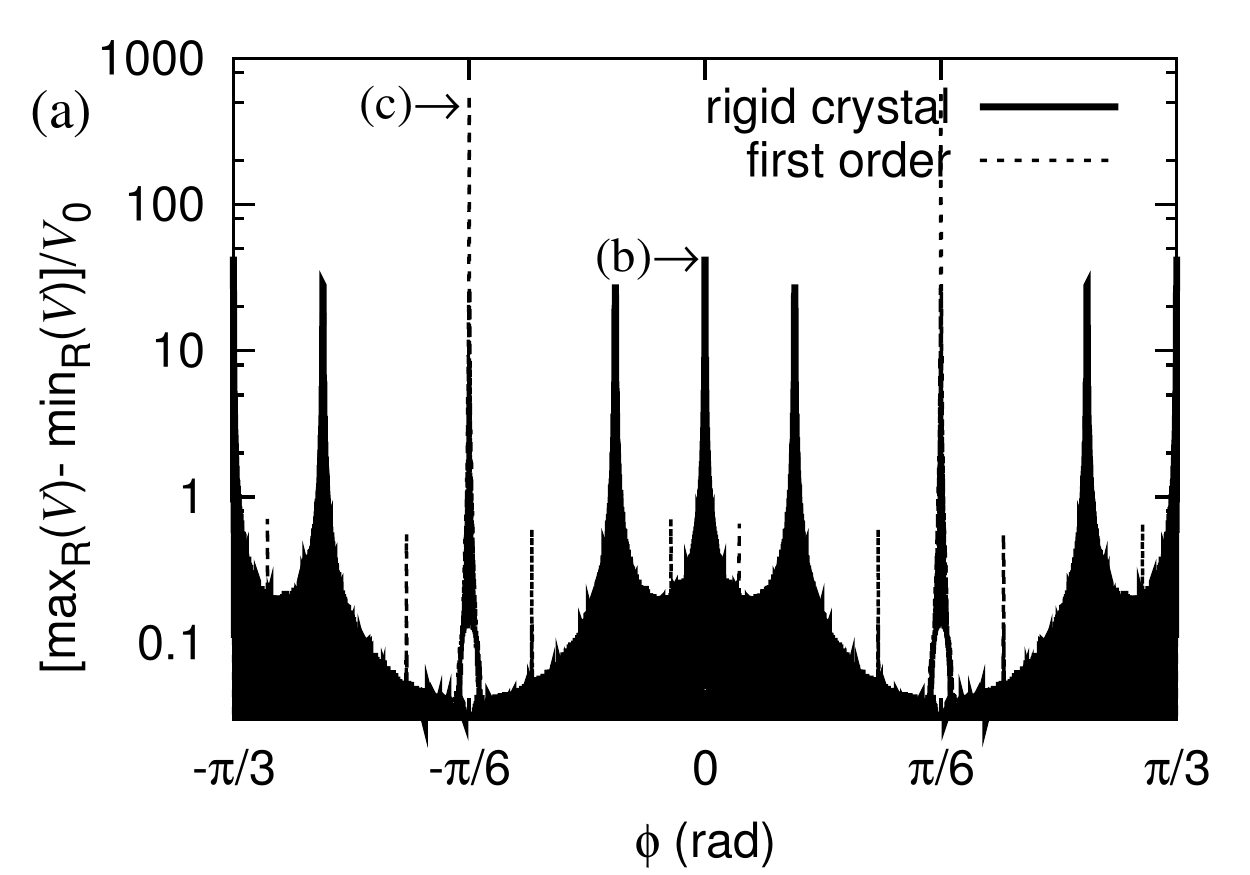}
\vskip\medskipamount
~~\includegraphics[width=8.2cm]{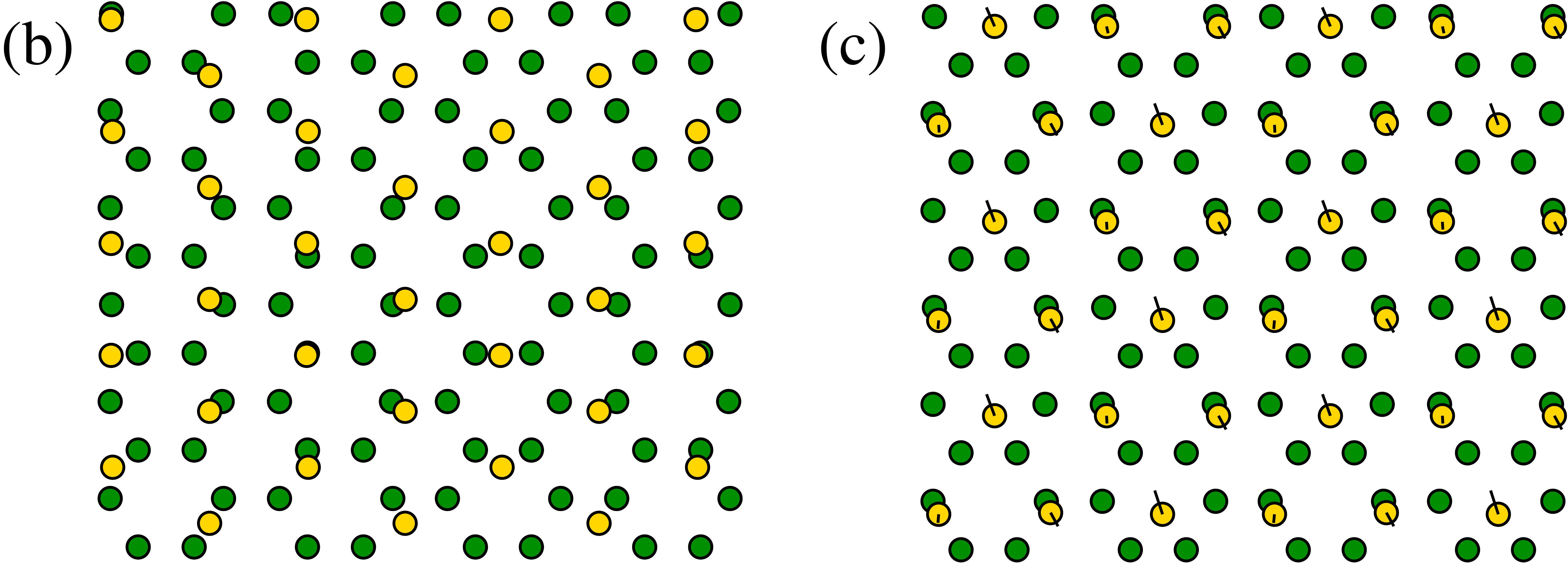}
\caption{
\label{fig:correctionsAu}
The corrugation (a) in the potential energy as a function of the orientation
and commensurate lattices (b) at $\phi=0,C=1$, (c) at $\phi=\pi/6,C=2$
for a perfectly triangular nanocrystal with $d=1000$ that is not rigid with $2V_0/(9 \kappa \alpha^2)=0.003$ and $b/\alpha=4/3$, which is a commensurate configuration for the first-order corrections,
This plot is similar to Fig.~\ref{fig:corrections}.
The additional peaks that appear because the crystal is not rigid are the solutions of Eqs.~(\ref{eq:denominator3}) -- (\ref{eq:denominator6}).
The orientation shown in (c) is usually strongly incommensurate, even though the symmetry axes area clearly lined up,
because the potential energy of the interface is not strongly corrugated.
Due to the matching lattice parameters, the first-order corrections contribute $O(d^2)$ to the corrugation.
For rigid crystals this relative orientation is incommensurate for almost all values of $b/\alpha$.  Even for $4/3$, which looks commensurate, a rigid crystal does not have a corrugation that increases with the size.
}
\end{figure}

For some lattice parameters, the rigid crystal has only two friction branches, but a third branch appears when lattice deformations are taken into account.
This is shown for a perfectly triangular non-rigid nanocrystal with $b/\alpha=4/3$ in Fig.~\ref{fig:correctionsAu}.
The first-order peaks appear at the solutions of Eqs.~(\ref{eq:denominator3}) -- (\ref{eq:denominator6}), and are generated by both rigid-crystal contributions and first-order corrections.
However, for this combination of lattice parameters, Eqs.~(\ref{eq:denominator5}) and~(\ref{eq:denominator6}) are satisfied for $p=2,p'=-2,0,2$. 
As a result, $C=2$ is possible for the first-order correction terms, but not for the zeroth order.
Consequently, the small prefactor for the stiffness of the crystal is multiplied by a factor that can be as large as $d$.

For sufficiently small crystals with nearly commensurate lattice parameters, the commensurate orientation behaves as if $C=2$, and its friction grows as $d^2$.
For gold on graphite, with $b/\alpha \approx 1.354$ close to $4/3$, the corrections can easily become as large as the leading order terms.
It should however be noted that, as this effect is very sensitive to the lattice parameters, the surface reconstruction, however minimal, can still enhance or decrease the effect strongly.
The surface construction of Au(111) on graphite may indeed itsself be affected by the near match in lattice parameters, possibly leading to an even closer match of the reconstructed surface.
The FK model can be used to understand surface recontructions~\cite{FKreconstruction}, and the calculations of this paper can similarly be applied to get a handle on surface reconstructions in an interface.

\subsection{Order of commensurability}
The order of commensuratebility, as it is defined in this work, is a general quantity that can be used to characterize commensurability of crystalline interfaces.
The $\sin$ functions in the denominator in Eq.~(\ref{eq:pottrap2}) originate from summing over a line of regularly spaced atoms, which must always produce factors of the type shown in Eq.~(\ref{eq:sinratio}).
Such sums occur in the potential energy of any rigid regular crystalline contact layer.
Thus, the analysis described here in terms of its derivatives can be used to identify and classify commensurate configurations for any regular nanocrystal on any regular substrate.
In the Appendix, the expression for the potential energy of a rectangular crystal on a rectangular substrate is worked out as an example.

For a two-dimensional contact area, it is not possible to have $C$ larger than 2, because the total potential energy can never grow faster than linearly with the contact area.
Correspondingly, in one dimension, $C$ cannot exceed unity.

As illustrated by the case of gold on graphite, configurations close to commensurate
still show some of the commensurate behavior for finite size crystals.
It is therefore worthwhile to consider a quantity similar to $C$, a finite-size order of commensurability,
\begin{align}
{\mathcal C} = \frac{2 \ln \left[ \max_\vec{R}\left(\frac{V}{V_0}\right)- \min_\vec{R}\left(\frac{V}{V_0}\right)\right]}{\ln {N}}~.
\end{align}
For simplicity, or if $V$ is not well-known, a simple periodic sinusoidal function can be used in place of $V$.
In the limit of $N\rightarrow\infty$, ${\mathcal C}$ converges to the order of commensurability $C$.
As it is impossible to define lattice orientation for a single atom, it diverges for $d=0$.
For finite-size crystals, it can be used as a measure of the commensurability of the contact.
In Ref.~\cite{marco} another measure for commensurability was defined
for determining to what extent weakly-coupled atoms line up with the substrate in an incommensurate contact.
This measure, however, is not suitable for determining the commensurability of two (nearly) rigid lattices.

\section{Conclusions\label{sec:conclusions}}

In this work, the (in)commensurability of finite-size crystalline interfaces was investigated, particularly for crystals with triangular symmetry on triangular or hexagonal substrates such as gold on graphite.
The crystals were assumed to be nearly rigid, with the stiffness being used as an expansion parameter.
A simple method was developed for determining the commensurate configurations, through quantifying the commensurability by the scaling of the potential corrugation of the crystal with the diameter of the contact area at constant orientation and lattice parameters.
This method can be applied not only to the two geometries discussed in this paper (triangular/hexagonal and rectangular/rectangular), but to any regular lattice on a regular substrate.

It was found that, due to the two-dimensional nature of the contact layer, and associated two parameters for commensurability (ratio of lattice parameters and relative orientation), two different types of commensurate contacts can exist for the same materials.
If the relative orientation is commensurate, but the lattice parameters are not, the potential corrugation scales with the diameter of the contact area, while if both are commensurate, it scales with the contact area.
Consequently, for some combinations of materials three different friction branches can appear for the same crystal shape and materials, scaling with the contact area as $A^0, A^{1/2}$, and $A^1$ for perfect crystals [see Eq.~(\ref{eq:scaling})].
When the crystals have irregular shapes, the  corrugation and friction  scale with $A^{1/4}, A^{3/4}, A^{1}$ [see Eq.~(\ref{eq:scalingimperfect})].

Nontrivial commensurate configurations were identified in particular for the triangular on triangular on hexagonal geometry.
These commensurate orientations exist if the lattice parameter the nanocrystal is sufficiently high compared to the lattice parameter of the substrate.
They are also a direct result of the two-dimensional nature of the contact layer.

It was found that lattice deformations produce a number of additional commensurate configurations that would be incommensurate for a rigid crystal.
Interestingly, the prototype geometry of gold on graphite is very close to the parameters which produce such a friction multiplicity.
It might therefore be possible to detect all three friction branches with different commensurability in friction experiments of gold nanocrystal on graphite, as well as nontrivial commensurate angles.

\begin{acknowledgments}

This work was financially supported by a Veni grant of Netherlands Organisation for Scientific Research (NWO) and by an Unga Forskare grant from the Swedish Research Council.
The author would also like to thank A.~Fasolino, D.~Dietzel, and A.~Schirmeisen for discussions, and O.~Manyuhina for carefully reading the manuscript.

\end{acknowledgments}

\appendix

\section{Rectangular lattices}

\begin{figure}
\includegraphics[width=7.8cm]{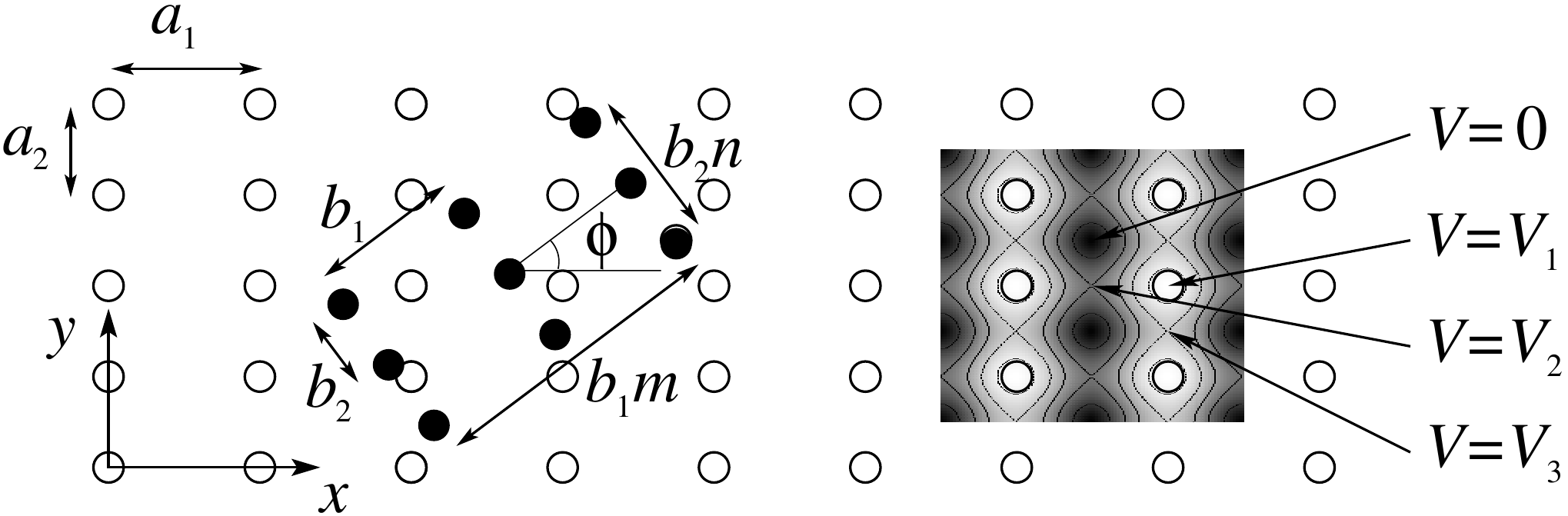}
\caption{
A top view of a general rectangular lattice (open circles) and contact layer (solid circles) with mismatch angle $\phi$, lattice parameters $a_1,a_2,b_1,b_2$, contact layer size $m,n$, and the potential energy of a contact layer atom on the substrate.
If a contact layer atom lies on top of a substrate atom, its potential energy is $V_1$.
If it lies directly between an atom and its nearest neighbor in the $x$ or $y$ direction, it has potential energy $V_2$ or $V_3$ respectively.
If it lies in the center of a rectangle, at equal distance from four substrate atoms, without loss of generality, we may set the potential energy to 0.
The origin of the coordinate system is chosen to lie on top of a substrate atom.
\label{fig:rectangular}
}
\end{figure}

The identification and classification of commensurate(-like) configurations described in this paper can easily be repeated for the general rectangular lattices, as described for instance in Ref.~\cite{onssquareflakes}.
A general potential energy of a single atom, $V_\mathrm{A}$ is given by
\begin{align}
\lefteqn{V_\mathrm{A}(X,Y)
 =  \frac{V_2+V_3}{2}
}&
\nonumber \\
&\phantom{=}
\null
+ \frac{V_1-V_2+V_3}{4} \cos \frac{2 \pi X}{a_1}
+ \frac{V_1+V_2-V_3}{4} \cos \frac{2 \pi Y}{a_2}
\nonumber \\
&\phantom{=}
\null
+ \frac{V_1-V_2-V_3}{4} \cos \frac{ 2 \pi X}{a_1} \cos \frac{ 2 \pi Y}{a_2}~,
\label{eq:VA}
\end{align}
where the geometry and associated parameters are defined in Fig.~\ref{fig:rectangular}.

The total potential energy of a rigid rectangular nanocrystal on the substrate can be calculated using the same approach as described in Sec.~\ref{sec:maths}, and is found to be
\begin{widetext}
\begin{align}
V(\vec{R},\phi) & = m n \frac{V_2 + V_3}{2} + {\mathcal V}_2 + {\mathcal V}_3 + {\mathcal V}_4~,\\
{\mathcal V}_2 &=
\frac{V_1-V_{2}+V_{3}}{4}
\frac{
\sin\left(\pi (m+1) \frac{b_1}{a_1}\cos{\phi}\right)
\sin\left(\pi (n+1) \frac{b_2}{a_1}\sin{\phi}\right)
}{
\sin\left(\pi \frac{b_1}{a_1}\cos{\phi}\right)
\sin\left(\pi \frac{b_2}{a_1}\sin{\phi}\right)
}
\cos\frac{2\pi X}{a_1}
~,\\
{\mathcal V}_3 &=  \frac{V_1-V_{3}+V_{2}}{4}
\frac{
\sin\left(\pi (m+1) \frac{b_1}{a_2}\sin{\phi}\right)
\sin\left(\pi (n+1) \frac{b_2}{a_2}\cos{\phi}\right)
}{
\sin\left(\pi \frac{b_1}{a_2}\sin{\phi}\right)
\sin\left(\pi \frac{b_2}{a_2}\cos{\phi}\right)
}
\cos\frac{2\pi Y}{a_2}
~,\\
{\mathcal V}_4 &= \frac{V_1-V_2-V_3}{8} \sum_{j=0,1}
\cos\left[2 \pi \left(\frac{X}{a_1} + (-1)^j\frac{Y}{a_2}\right)\right]
\nonumber\\
&\phantom{= nnnn}
\times
\frac{
\sin\left[\pi (m+1) \left(\frac{b_1}{a_1}\cos{\phi} + (-1)^j \frac{b_1}{a_2}\sin{\phi}\right)\right]
\sin\left[\pi (n+1) \left(\frac{b_2}{a_1}\sin{\phi} - (-1)^j \frac{b_2}{a_2}\cos{\phi}\right)\right]
}{
\sin\left[\pi \left(\frac{b_1}{a_1}\cos{\phi} + (-1)^j \frac{b_1}{a_2}\sin{\phi}\right)\right]
\sin\left[\pi \left(\frac{b_2}{a_1}\sin{\phi} - (-1)^j \frac{b_2}{a_2}\cos{\phi}\right)\right]
}
~,
\end{align}
\end{widetext}
where ${\mathcal V}_2, {\mathcal V}_3, {\mathcal V}_4$ originate from the second, third, and fourth term on the right-hand side of Eq.~(\ref{eq:VA}).
Similarly, the higher-order corrections for lattice deformations can also be obtained along similar lines as described in Sec.~\ref{sec:firstordercorrections}.
This yields terms with higher perdiodicity, i.e., a term obtained from the derivative of ${\mathcal V}_2$ which looks very similar, except that $a_1$ is replaced by $2 a_1$.
Similarly, there is a term derived from ${\mathcal V}_3$ with $a_2$ replaced by $2a_2$, and two terms from ${\mathcal V}_4$ with both $a_1$ and $a_2$ replaced.

From these expressions it is easy to obtain the conditions for vanishing denominators, and thus for $C=0,1,2$.
The three terms ${\mathcal V}_2, {\mathcal V}_3, {\mathcal V}_4$ each produce a different set of conditions for commensurability at different angles and ratios of lattice parameters.
For ${\mathcal V}_2$, one finds that $C=1$ commensurability occurs if one of the following conditions is met,
\begin{align}
\label{eq:denominatorsquare1}
\frac{b_1}{a_1} \cos\phi_{\mathrm{c.}}  = p~,\\
\label{eq:denominatorsquare2}
\frac{b_2}{a_1} \sin\phi_{\mathrm{c.}}  = p'~,
\end{align}
for some $p,p'\in\mathbb{Z}$.
Both of these conditions can be met simulataneously, leading to $C=2$ commensurability, if
\begin{align}
\label{eq:denominatorsquarex1}
\left(\frac{a_1 p}{b_1}\right)^2 + \left(\frac{a_1 p'}{b_2}\right)^2 = 1~.
\end{align}

For ${\mathcal V}_3$, one finds,
\begin{align}
\label{eq:denominatorsquare3}
\frac{b_2}{a_2} \cos\phi_{\mathrm{c.}}  = p~,\\
\label{eq:denominatorsquare4}
\frac{b_1}{a_2} \sin\phi_{\mathrm{c.}}  = p'~.
\end{align}
Commensurability with $C=2$ occurs if
\begin{align}
\label{eq:denominatorsquarex2}
\left(\frac{a_2 p}{b_2}\right)^2 + \left(\frac{a_2 p'}{b_1}\right)^2 = 1~.
\end{align}

Finally, for ${\mathcal V}_4$, one finds the conditions for $C=1$,
\begin{align}
\label{eq:denominatorsquare5}
\frac{b_1}{a_1}\cos\phi_{\mathrm{c.}} + (-1)^j \frac{b_1}{a_2}\sin\phi_{\mathrm{c.}} &= p~,\\
\label{eq:denominatorsquare6}
\frac{b_2}{a_1}\sin\phi_{\mathrm{c.}} - (-1)^j \frac{b_2}{a_2}\cos\phi_{\mathrm{c.}} &= p'~,
\end{align}
and for $C=2$, the lattice parameters must satisfy both of these equations at the same time.

The first-order correction for stiff, but not rigid, crystals, adds additional commensurate orientations, which can be found by performing the replacements of $a_1$ and $a_2$ described above.

As each of ${\mathcal V}_2, {\mathcal V}_3, {\mathcal V}_4$ has a different prefactor that depends on the energy barriers of the potential, the different terms do not, in general, cancel each other out.
It is interesting to note that often for square lattices it is assumed that $V_2=V_3=V_1/2$.
In this case, ${\mathcal V}_4$ vanishes and the commensurate configurations associated with it disappear, leading to lower friction.


\begin{thebibliography}{31}
\expandafter\ifx\csname natexlab\endcsname\relax\def\natexlab#1{#1}\fi
\expandafter\ifx\csname bibnamefont\endcsname\relax
  \def\bibnamefont#1{#1}\fi
\expandafter\ifx\csname bibfnamefont\endcsname\relax
  \def\bibfnamefont#1{#1}\fi
\expandafter\ifx\csname citenamefont\endcsname\relax
  \def\citenamefont#1{#1}\fi
\expandafter\ifx\csname url\endcsname\relax
  \def\url#1{\texttt{#1}}\fi
\expandafter\ifx\csname urlprefix\endcsname\relax\def\urlprefix{URL }\fi
\providecommand{\bibinfo}[2]{#2}
\providecommand{\eprint}[2][]{\url{#2}}

\bibitem[{\citenamefont{Peyrard and Aubry}(1983)}]{vanishingstaticfriction}
\bibinfo{author}{\bibfnamefont{M.}~\bibnamefont{Peyrard}} \bibnamefont{and}
  \bibinfo{author}{\bibfnamefont{S.}~\bibnamefont{Aubry}},
  \bibinfo{journal}{J.~Phys.~C} \textbf{\bibinfo{volume}{16}},
  \bibinfo{pages}{1593} (\bibinfo{year}{1983}).

\bibitem[{\citenamefont{Aubry and de~Seze}(1985)}]{vanishingfriction2}
\bibinfo{author}{\bibfnamefont{S.}~\bibnamefont{Aubry}} \bibnamefont{and}
  \bibinfo{author}{\bibfnamefont{L.}~\bibnamefont{de~Seze}}, in
  \emph{\bibinfo{booktitle}{Festk\"orperprobleme 25}}, edited by
  \bibinfo{editor}{\bibfnamefont{P.}~\bibnamefont{Grosse}}
  (\bibinfo{publisher}{Springer Berlin / Heidelberg}, \bibinfo{year}{1985}),
  vol.~\bibinfo{volume}{25} of \emph{\bibinfo{series}{Advances in Solid State
  Physics}}, pp. \bibinfo{pages}{59--69}.

\bibitem[{\citenamefont{Shinjo and Hirano}(1993)}]{Shinjo}
\bibinfo{author}{\bibfnamefont{K.}~\bibnamefont{Shinjo}} \bibnamefont{and}
  \bibinfo{author}{\bibfnamefont{M.}~\bibnamefont{Hirano}},
  \bibinfo{journal}{Surf.~Sci.} \textbf{\bibinfo{volume}{283}},
  \bibinfo{pages}{473} (\bibinfo{year}{1993}).

\bibitem[{\citenamefont{Consoli et~al.}(2000)\citenamefont{Consoli, Knops, and
  Fasolino}}]{fkphononconsoli}
\bibinfo{author}{\bibfnamefont{L.}~\bibnamefont{Consoli}},
  \bibinfo{author}{\bibfnamefont{H.~J.~F.} \bibnamefont{Knops}},
  \bibnamefont{and} \bibinfo{author}{\bibfnamefont{A.}~\bibnamefont{Fasolino}},
  \bibinfo{journal}{Phys. Rev. Lett.} \textbf{\bibinfo{volume}{85}},
  \bibinfo{pages}{302} (\bibinfo{year}{2000}).

\bibitem[{\citenamefont{Dienwiebel et~al.}(2004)\citenamefont{Dienwiebel,
  Verhoeven, Pradeep, Frenken, Heimberg, and Zandbergen}}]{Dienwiebel2004}
\bibinfo{author}{\bibfnamefont{M.}~\bibnamefont{Dienwiebel}},
  \bibinfo{author}{\bibfnamefont{G.~S.} \bibnamefont{Verhoeven}},
  \bibinfo{author}{\bibfnamefont{N.}~\bibnamefont{Pradeep}},
  \bibinfo{author}{\bibfnamefont{J.~W.~M.} \bibnamefont{Frenken}},
  \bibinfo{author}{\bibfnamefont{J.~A.} \bibnamefont{Heimberg}},
  \bibnamefont{and} \bibinfo{author}{\bibfnamefont{H.~W.}
  \bibnamefont{Zandbergen}}, \bibinfo{journal}{Phys.~Rev.~Lett.}
  \textbf{\bibinfo{volume}{92}}, \bibinfo{pages}{126101}
  (\bibinfo{year}{2004}).

\bibitem[{\citenamefont{Filippov et~al.}(2008)\citenamefont{Filippov,
  Dienwiebel, Frenken, Klafter, and Urbakh}}]{torqueandtwist}
\bibinfo{author}{\bibfnamefont{A.~E.} \bibnamefont{Filippov}},
  \bibinfo{author}{\bibfnamefont{M.}~\bibnamefont{Dienwiebel}},
  \bibinfo{author}{\bibfnamefont{J.~W.~M.} \bibnamefont{Frenken}},
  \bibinfo{author}{\bibfnamefont{J.}~\bibnamefont{Klafter}}, \bibnamefont{and}
  \bibinfo{author}{\bibfnamefont{M.}~\bibnamefont{Urbakh}},
  \bibinfo{journal}{Phys. Rev. Lett.} \textbf{\bibinfo{volume}{100}},
  \bibinfo{pages}{046102} (\bibinfo{year}{2008}).

\bibitem[{\citenamefont{Dietzel et~al.}(2008)\citenamefont{Dietzel, Ritter,
  M\"onninghoff, Fuchs, Schirmeisen, and Schwarz}}]{Dietzel2008}
\bibinfo{author}{\bibfnamefont{D.}~\bibnamefont{Dietzel}},
  \bibinfo{author}{\bibfnamefont{C.}~\bibnamefont{Ritter}},
  \bibinfo{author}{\bibfnamefont{T.}~\bibnamefont{M\"onninghoff}},
  \bibinfo{author}{\bibfnamefont{H.}~\bibnamefont{Fuchs}},
  \bibinfo{author}{\bibfnamefont{A.}~\bibnamefont{Schirmeisen}},
  \bibnamefont{and} \bibinfo{author}{\bibfnamefont{U.~D.}
  \bibnamefont{Schwarz}}, \bibinfo{journal}{Phys.~Rev.~Lett.}
  \textbf{\bibinfo{volume}{101}}, \bibinfo{pages}{125505}
  (\bibinfo{year}{2008}).

\bibitem[{\citenamefont{Dietzel et~al.}(2010)\citenamefont{Dietzel,
  M\"onninghoff, Herding, Feldmann, Fuchs, Stegemann, Ritter, Schwarz, and
  Schirmeisen}}]{dietzelprb}
\bibinfo{author}{\bibfnamefont{D.}~\bibnamefont{Dietzel}},
  \bibinfo{author}{\bibfnamefont{T.}~\bibnamefont{M\"onninghoff}},
  \bibinfo{author}{\bibfnamefont{C.}~\bibnamefont{Herding}},
  \bibinfo{author}{\bibfnamefont{M.}~\bibnamefont{Feldmann}},
  \bibinfo{author}{\bibfnamefont{H.}~\bibnamefont{Fuchs}},
  \bibinfo{author}{\bibfnamefont{B.}~\bibnamefont{Stegemann}},
  \bibinfo{author}{\bibfnamefont{C.}~\bibnamefont{Ritter}},
  \bibinfo{author}{\bibfnamefont{U.~D.} \bibnamefont{Schwarz}},
  \bibnamefont{and}
  \bibinfo{author}{\bibfnamefont{A.}~\bibnamefont{Schirmeisen}},
  \bibinfo{journal}{Phys.~Rev.~B} \textbf{\bibinfo{volume}{82}},
  \bibinfo{pages}{035401} (\bibinfo{year}{2010}).

\bibitem[{\citenamefont{{de Wijn} et~al.}(2010)\citenamefont{{de Wijn}, Fusco,
  and Fasolino}}]{onsgraphiteflakes}
\bibinfo{author}{\bibfnamefont{A.~S.} \bibnamefont{{de Wijn}}},
  \bibinfo{author}{\bibfnamefont{C.}~\bibnamefont{Fusco}}, \bibnamefont{and}
  \bibinfo{author}{\bibfnamefont{A.}~\bibnamefont{Fasolino}},
  \bibinfo{journal}{Phys.~Rev.~E} \textbf{\bibinfo{volume}{81}},
  \bibinfo{pages}{046105} (\bibinfo{year}{2010}).

\bibitem[{\citenamefont{{de Wijn} and Fasolino}(2010)}]{onssquareflakes}
\bibinfo{author}{\bibfnamefont{A.~S.} \bibnamefont{{de Wijn}}}
  \bibnamefont{and} \bibinfo{author}{\bibfnamefont{A.}~\bibnamefont{Fasolino}},
  \bibinfo{journal}{Tribol.~Lett.} \textbf{\bibinfo{volume}{39}},
  \bibinfo{pages}{91} (\bibinfo{year}{2010}).

\bibitem[{\citenamefont{Dietzel et~al.}(2012)\citenamefont{Dietzel, Feldmann,
  Fuchs, Schwarz, and Schirmeisen}}]{dietzeltbp}
\bibinfo{author}{\bibfnamefont{D.}~\bibnamefont{Dietzel}},
  \bibinfo{author}{\bibfnamefont{M.}~\bibnamefont{Feldmann}},
  \bibinfo{author}{\bibfnamefont{H.}~\bibnamefont{Fuchs}},
  \bibinfo{author}{\bibfnamefont{U.~D.} \bibnamefont{Schwarz}},
  \bibnamefont{and}
  \bibinfo{author}{\bibfnamefont{A.}~\bibnamefont{Schirmeisen}}
  (\bibinfo{year}{2012}), \bibinfo{note}{private communication}.

\bibitem[{\citenamefont{Guerra et~al.}(2010)\citenamefont{Guerra, Tartaglino,
  Vanossi, and Tosatti}}]{vanossinature2010}
\bibinfo{author}{\bibfnamefont{R.}~\bibnamefont{Guerra}},
  \bibinfo{author}{\bibfnamefont{U.}~\bibnamefont{Tartaglino}},
  \bibinfo{author}{\bibfnamefont{A.}~\bibnamefont{Vanossi}}, \bibnamefont{and}
  \bibinfo{author}{\bibfnamefont{E.}~\bibnamefont{Tosatti}},
  \bibinfo{journal}{Nature Materials} \textbf{\bibinfo{volume}{9}},
  \bibinfo{pages}{634} (\bibinfo{year}{2010}).

\bibitem[{\citenamefont{Schirmeisen and Schwarz}(2009)}]{Schirmeisen2009}
\bibinfo{author}{\bibfnamefont{A.}~\bibnamefont{Schirmeisen}} \bibnamefont{and}
  \bibinfo{author}{\bibfnamefont{U.~D.} \bibnamefont{Schwarz}},
  \bibinfo{journal}{ChemPhysChem} \textbf{\bibinfo{volume}{10}},
  \bibinfo{pages}{2373} (\bibinfo{year}{2009}).

\bibitem[{\citenamefont{Jensen et~al.}(2004)\citenamefont{Jensen, Blase, and
  Ordejón}}]{goldgraphitediffusionbarrier}
\bibinfo{author}{\bibfnamefont{P.}~\bibnamefont{Jensen}},
  \bibinfo{author}{\bibfnamefont{X.}~\bibnamefont{Blase}}, \bibnamefont{and}
  \bibinfo{author}{\bibfnamefont{P.}~\bibnamefont{Ordejón}},
  \bibinfo{journal}{Surface Science} \textbf{\bibinfo{volume}{564}},
  \bibinfo{pages}{173} (\bibinfo{year}{2004}).

\bibitem[{\citenamefont{Harten et~al.}(1985)\citenamefont{Harten, Lahee,
  Toennies, and W\"oll}}]{aureconstruction}
\bibinfo{author}{\bibfnamefont{U.}~\bibnamefont{Harten}},
  \bibinfo{author}{\bibfnamefont{A.~M.} \bibnamefont{Lahee}},
  \bibinfo{author}{\bibfnamefont{J.~P.} \bibnamefont{Toennies}},
  \bibnamefont{and} \bibinfo{author}{\bibfnamefont{C.}~\bibnamefont{W\"oll}},
  \bibinfo{journal}{Phys. Rev. Lett.} \textbf{\bibinfo{volume}{54}},
  \bibinfo{pages}{2619} (\bibinfo{year}{1985}).

\bibitem[{\citenamefont{Persson and Tosatti}(1999)}]{perssontosatti}
\bibinfo{author}{\bibfnamefont{B.}~\bibnamefont{Persson}} \bibnamefont{and}
  \bibinfo{author}{\bibfnamefont{E.}~\bibnamefont{Tosatti}},
  \bibinfo{journal}{Solid State Communications} \textbf{\bibinfo{volume}{109}},
  \bibinfo{pages}{739} (\bibinfo{year}{1999}).

\bibitem[{\citenamefont{Sokoloff}(2001)}]{Sokoloff2000}
\bibinfo{author}{\bibfnamefont{J.~B.} \bibnamefont{Sokoloff}},
  \bibinfo{journal}{Phys. Rev. Lett.} \textbf{\bibinfo{volume}{86}},
  \bibinfo{pages}{3312} (\bibinfo{year}{2001}).

\bibitem[{\citenamefont{M\"user}(2004)}]{mueserstructurallubricity}
\bibinfo{author}{\bibfnamefont{M.~H.} \bibnamefont{M\"user}},
  \bibinfo{journal}{Europhysics Letters} \textbf{\bibinfo{volume}{66}},
  \bibinfo{pages}{97} (\bibinfo{year}{2004}).

\bibitem[{\citenamefont{Rappe et~al.}(1992)\citenamefont{Rappe, Casewit,
  Colwell, Goddard, and Skiff}}]{goldforcefield}
\bibinfo{author}{\bibfnamefont{A.~K.} \bibnamefont{Rappe}},
  \bibinfo{author}{\bibfnamefont{C.~J.} \bibnamefont{Casewit}},
  \bibinfo{author}{\bibfnamefont{K.~S.} \bibnamefont{Colwell}},
  \bibinfo{author}{\bibfnamefont{W.~A.} \bibnamefont{Goddard}},
  \bibnamefont{and} \bibinfo{author}{\bibfnamefont{W.~M.} \bibnamefont{Skiff}},
  \bibinfo{journal}{Journal of the American Chemical Society}
  \textbf{\bibinfo{volume}{114}}, \bibinfo{pages}{10024}
  (\bibinfo{year}{1992}).

\bibitem[{\citenamefont{Manini and Braun}(2011)}]{Manini2011}
\bibinfo{author}{\bibfnamefont{N.}~\bibnamefont{Manini}} \bibnamefont{and}
  \bibinfo{author}{\bibfnamefont{O.~M.} \bibnamefont{Braun}},
  \bibinfo{journal}{Phys.~Lett.~A} \textbf{\bibinfo{volume}{375}},
  \bibinfo{pages}{2946} (\bibinfo{year}{2011}).

\bibitem[{\citenamefont{Ritter et~al.}(2005)\citenamefont{Ritter, Heyde,
  Stegemann, Rademann, and Schwarz}}]{oldantimony}
\bibinfo{author}{\bibfnamefont{C.}~\bibnamefont{Ritter}},
  \bibinfo{author}{\bibfnamefont{M.}~\bibnamefont{Heyde}},
  \bibinfo{author}{\bibfnamefont{B.}~\bibnamefont{Stegemann}},
  \bibinfo{author}{\bibfnamefont{K.}~\bibnamefont{Rademann}}, \bibnamefont{and}
  \bibinfo{author}{\bibfnamefont{U.~D.} \bibnamefont{Schwarz}},
  \bibinfo{journal}{Phys.~Rev.~B} \textbf{\bibinfo{volume}{71}},
  \bibinfo{pages}{085405} (\bibinfo{year}{2005}).

\bibitem[{\citenamefont{Frenkel and Kontorova}(1938)}]{frenkelkontorova}
\bibinfo{author}{\bibfnamefont{Y.~I.} \bibnamefont{Frenkel}} \bibnamefont{and}
  \bibinfo{author}{\bibfnamefont{T.~A.} \bibnamefont{Kontorova}},
  \bibinfo{journal}{Zh.~Eksp.~Teor.~Fiz.} \textbf{\bibinfo{volume}{8}},
  \bibinfo{pages}{89} (\bibinfo{year}{1938}).

\bibitem[{\citenamefont{Reguzzoni et~al.}(2010)\citenamefont{Reguzzoni,
  Ferrario, Zapperi, and Rihi}}]{marcoPNAS}
\bibinfo{author}{\bibfnamefont{M.}~\bibnamefont{Reguzzoni}},
  \bibinfo{author}{\bibfnamefont{M.}~\bibnamefont{Ferrario}},
  \bibinfo{author}{\bibfnamefont{S.}~\bibnamefont{Zapperi}}, \bibnamefont{and}
  \bibinfo{author}{\bibfnamefont{M.~C.} \bibnamefont{Rihi}},
  \bibinfo{journal}{Proc. Nat. Acad. Sci. USA} \textbf{\bibinfo{volume}{107}},
  \bibinfo{pages}{1311} (\bibinfo{year}{2010}).

\bibitem[{\citenamefont{Reguzzoni and Righi}(2012)}]{marco}
\bibinfo{author}{\bibfnamefont{M.}~\bibnamefont{Reguzzoni}} \bibnamefont{and}
  \bibinfo{author}{\bibfnamefont{M.~C.} \bibnamefont{Righi}},
  \bibinfo{journal}{Phys.~Rev.~B} \textbf{\bibinfo{volume}{85}},
  \bibinfo{pages}{201412(R)} (\bibinfo{year}{2012}).

\bibitem[{\citenamefont{van Erp et~al.}(1999)\citenamefont{van Erp, Fasolino,
  Radulescu, and Janssen}}]{quasiperiodicFK}
\bibinfo{author}{\bibfnamefont{T.~S.} \bibnamefont{van Erp}},
  \bibinfo{author}{\bibfnamefont{A.}~\bibnamefont{Fasolino}},
  \bibinfo{author}{\bibfnamefont{O.}~\bibnamefont{Radulescu}},
  \bibnamefont{and} \bibinfo{author}{\bibfnamefont{T.}~\bibnamefont{Janssen}},
  \bibinfo{journal}{Phys. Rev. B} \textbf{\bibinfo{volume}{60}},
  \bibinfo{pages}{6522} (\bibinfo{year}{1999}).

\bibitem[{\citenamefont{Lomdahl and Srolovitz}(1986)}]{2dFK-1}
\bibinfo{author}{\bibfnamefont{P.~S.} \bibnamefont{Lomdahl}} \bibnamefont{and}
  \bibinfo{author}{\bibfnamefont{D.~J.} \bibnamefont{Srolovitz}},
  \bibinfo{journal}{Phys.~Rev.~Lett.} \textbf{\bibinfo{volume}{57}},
  \bibinfo{pages}{2702} (\bibinfo{year}{1986}).

\bibitem[{\citenamefont{Braun et~al.}(1997)\citenamefont{Braun, Dauxois, Paliy,
  and Peyrard}}]{2dFK-2}
\bibinfo{author}{\bibfnamefont{O.~M.} \bibnamefont{Braun}},
  \bibinfo{author}{\bibfnamefont{T.}~\bibnamefont{Dauxois}},
  \bibinfo{author}{\bibfnamefont{M.~V.} \bibnamefont{Paliy}}, \bibnamefont{and}
  \bibinfo{author}{\bibfnamefont{M.}~\bibnamefont{Peyrard}},
  \bibinfo{journal}{Phys.~Rev.~E} \textbf{\bibinfo{volume}{55}},
  \bibinfo{pages}{3598} (\bibinfo{year}{1997}).

\bibitem[{\citenamefont{Teki\'c et~al.}(2005)\citenamefont{Teki\'c, Braun, and
  Hu}}]{2dFK-3}
\bibinfo{author}{\bibfnamefont{J.}~\bibnamefont{Teki\'c}},
  \bibinfo{author}{\bibfnamefont{O.~M.} \bibnamefont{Braun}}, \bibnamefont{and}
  \bibinfo{author}{\bibfnamefont{B.}~\bibnamefont{Hu}},
  \bibinfo{journal}{Phys.~Rev.~E} \textbf{\bibinfo{volume}{71}},
  \bibinfo{pages}{026104} (\bibinfo{year}{2005}).

\bibitem[{\citenamefont{Bohlein et~al.}(2012)\citenamefont{Bohlein, Mikhael,
  and Bechinger}}]{colloidFK}
\bibinfo{author}{\bibfnamefont{T.}~\bibnamefont{Bohlein}},
  \bibinfo{author}{\bibfnamefont{J.}~\bibnamefont{Mikhael}}, \bibnamefont{and}
  \bibinfo{author}{\bibfnamefont{C.}~\bibnamefont{Bechinger}},
  \bibinfo{journal}{Nat.~Mater.} \textbf{\bibinfo{volume}{11}},
  \bibinfo{pages}{126} (\bibinfo{year}{2012}).

\bibitem[{\citenamefont{M\"user et~al.}(2001)\citenamefont{M\"user, Wenning,
  and Robbins}}]{mueserprl}
\bibinfo{author}{\bibfnamefont{M.~H.} \bibnamefont{M\"user}},
  \bibinfo{author}{\bibfnamefont{L.}~\bibnamefont{Wenning}}, \bibnamefont{and}
  \bibinfo{author}{\bibfnamefont{M.~O.} \bibnamefont{Robbins}},
  \bibinfo{journal}{Phys.~Rev.~Lett.} \textbf{\bibinfo{volume}{86}},
  \bibinfo{pages}{1295} (\bibinfo{year}{2001}).

\bibitem[{\citenamefont{Mansfield and Needs}(1990)}]{FKreconstruction}
\bibinfo{author}{\bibfnamefont{M.}~\bibnamefont{Mansfield}} \bibnamefont{and}
  \bibinfo{author}{\bibfnamefont{R.~J.} \bibnamefont{Needs}},
  \bibinfo{journal}{J.~Phys.:\ Cond.\ Mat} \textbf{\bibinfo{volume}{2}},
  \bibinfo{pages}{2361} (\bibinfo{year}{1990}).

\end{thebibliography}
\end{document}